\documentclass{article}
\usepackage[utf8]{inputenc}
\usepackage{amsmath}
\usepackage{tabularx}
\usepackage[numbers]{natbib}
\usepackage{graphicx}

\title{Roadway Design Matters: Variation in Bicyclists' Psycho-Physiological Responses in Different Urban Roadway Designs}
\author{Xiang Guo, Arash Tavakoli, Erin Robartes,\\ Austin Angulo, T. Donna Chen, Arsalan Heydarian}
\date{Feb. 2022}

\begin{document}

\maketitle

\begin{abstract}
As a healthier and more sustainable way of mobility, cycling has been advocated by literature and policy. However, current trends in bicyclist crash fatalities suggest deficiencies in current roadway design in protecting these vulnerable road users. The lack of cycling data is a common challenge for studying bicyclists' safety, behavior, and comfort levels under different design contexts. To understand bicyclists' behavioral and physiological responses in an efficient and safe way, this study uses a bicycle simulator within an immersive virtual environment (IVE). Off-the-shelf sensors are utilized to evaluate bicyclists' cycling performance (speed and lane position) and physiological responses (eye tracking and heart rate (HR)). Participants bike in a simulated virtual environment modeled to scale from a real-world street with a shared bike lane (sharrow) to evaluate how introduction of a bike lane and a protected bike lane with pylons may impact perceptions of safety, as well as behavioral and psycho-physiological responses. Results from 50 participants show that the protected bike lane design received the highest perceived safety rating and exhibited the lowest average cycling speed. Furthermore, both the bike lane and the protected bike lane scenarios show a less dispersed gaze distribution than the as-built sharrow scenario, reflecting a higher gaze focus among bicyclists on the biking task in the bike lane and protected bike lane scenarios, compared to when bicyclists share right of way with vehicles. Additionally, heart rate change point results from the study suggest that creating dedicated zones for bicyclists (bike lanes or protected bike lanes) has the potential to reduce bicyclists' stress levels.
\end{abstract}

\section{Introduction} 
Bicycling, as a traditional means of mobility, has gained more popularity in recent years. Bicycle mode share is rising in response to issues in modern cities, such as increases in traffic jams, land use, energy consumption, air pollution, climate change, and physical inactivity \cite{flusche2012bicycling,rupi2019visual}. This trend has continued during the COVID-19 pandemic, as it is reported that bicycling levels have significantly increased in many countries despite lockdowns and travel restrictions \cite{buehler2021covid}. However, there has been an alarming increase in bicyclist fatalities over the last decade. The National Highway Traffic Safety Administration's report shows that in the United States, the number of bicyclists' fatalities has increased by more than 35\% since 2010 \cite{national2021fatality}. One potential reason for this increase is due to the auto-centric nature of much of US roadway design, which often overlooks the safety and comfort needs of vulnerable road users like bicyclists \cite{schultheiss2018historical}. Furthermore, current crash reports only focus on vehicle-bicyclist crashes that result in fatalities, and have very limited information on bicyclists' discomfort, behaviors, and responses in different roadway designs and contextual settings \cite{robartes2017effect}. The absence of applicable data has been recognized as a limiting factor for many transport and urban planning studies, especially for bicyclists \cite{willberg2021comparing}. Therefore, it is necessary to identify innovative approaches to understand how bicyclists' behavior, sense of safety, and comfort is affected under different roadway conditions during the design and planning phases. To achieve this, bicyclists' behavior and psycho-physiological state data should be explored and evaluated in different roadway settings. 

Various methods of collecting bicycle safety, risk, comfort and behavior data have been utilized in the past. These studies spanned across surveys, observational studies, naturalistic studies and simulation studies. For example, interviews and surveys ask participants about their behaviors and comfort in a certain design context either by imagination or after an actual bike ride.  However, the subjective response does not always reflect what the participant will do in a real-road setting and can suffer from hypothetical bias \cite{fitch2018relationship}. Observational studies can record realistic changes within the environment and bicyclists' responses in real-world conditions, but are unable to track bicyclists' physiological changes \cite{chidester2001pedestrian}. Naturalistic studies can further record bicyclists' responses in real-world conditions through different sensing modalities such as GPS, ECG, or mobile eye trackers \cite{rupi2019visual}. However, these studies have potential risks for participants as they may be placed in dangerous roadway settings (e.g. distraction in high traffic density area) \cite{stelling2018study}. Furthermore, in naturalistic studies, it is difficult to control many environmental factors that may impact the independent variables, especially for physiological and behavioral factors, which makes causal inferences especially difficult \cite{fitch2020psychological,teixeira2020does}. Experimental studies conducted through Immersive Virtual Environments (IVE) are an emerging approach that minimizes the hypothetical bias of subjective surveys while offering a controlled, low-risk, and immersive environment to evaluate the responses of bicyclists to different roadway designs and conditions. One of the main challenges in previous IVE simulation studies was the integration of human sensing techniques into the experiment. In IVE related literature, participants' physiological responses have been applied to evaluate different design alternatives for buildings \cite{francisco2018occupant}, hospitals \cite{chias20193d}, and other civil infrastructure systems \cite{awada2021integrated}. However, only a few recent studies have applied bicyclist physiological sensing in IVE simulators \cite{cobb2021bicyclists}, and a deeper understanding of bicyclists' psychological and physiological responses in different roadway design and conditions is still needed. Overall, we still have very a limited understanding of bicyclists' physiological responses in different roadway environments, especially in IVE studies. Some naturalistic studies have been conducted to evaluate bicyclist's behavior and physiological responses in different contextual settings \cite{guo_robartes_angulo_chen_heydarian_2021, mcneil2015influence,rupi2019visual,teixeira2020does}. These preliminary studies revealed that  psycho-physiological metrics (e.g., heart rate (HR), gaze variability, and skin conductance) are indicators of how participants' behaviors and perceptions may change in different contextual settings. 

By integrating a bicycle simulator with an IVE, this research aims to overcome some of the limitations found in previous research by conducting a repeated measures experiment to collect bicyclists' physiological responses (specifically, gaze variability and HR) in different urban roadway designs. With an IVE, we are able to control other roadway environmental factors, such as infrastructure design, vehicle traffic volume, traffic signal phase, vehicular speeds and gaps, vehicle types, lighting, and weather conditions to better quantify the relationships between bicyclists' behavior (e.g., speed, lane position) and physiological responses (e.g., gaze variability, HR). These measurements can be used as surrogate data to better understand how different types of roadway conditions and infrastructure designs result in higher rates of physiological stimulation. In this paper, we leverage human sensing tools (e.g., wearable devices) together with a bicycle simulator in IVE to specifically evaluate and model the bicyclists' psycho-physiological and behavioral responses. We consider three roadway design scenarios (shared bike lane [sharrow], standard dedicated bike lane, and protected bike lane with pylons) and evaluate bicyclists' HR, gaze measures, and speed within each environment. After performing extensive feature extractions on HR and gaze data, we leverage linear mixed effect models to compare bicyclists' responses across the simulated environments.

In this paper, we first provide detailed background on previous bicycle safety studies in naturalistic, simulated, and virtual reality (VR) environments. We then discuss the applicability of bicyclists' physiological responses and gaze measures in understanding their behaviors and physiological states, such as stress level and cognitive load. By providing the methodology of our experimental design, we dive into the details of the experiment. We then provide the results of bicyclist's HR and gaze variability together with bicycling performance within each environment with a linear mixed effect modeling approach. We conclude with a discussion on comparing different physiological responses and reasons behind the results.

\section{Literature Review}
This section is divided into literature reviews of different categories of bicyclists' behavior studies, including stated preference surveys (\ref{sec:preference}), observational studies (\ref{sec:observational}), naturalistic studies (\ref{sec:real_world}), and bicycle simulators (\ref{sec:simulators}). An overview of different measures of human psycho-physiology in experimental settings with a focus on bicyclists studies follows (\ref{sec:psycho_measure}).

\subsection{Stated Preferences Survey} \label{sec:preference}
Surveys have been widely used to study bicyclists' behavior, particularly when faced with a lack of real-world data. Surveys, when composed carefully, can efficiently assess large populations of bicyclists and have been used to study a wide variety of topics such as perceived safety and comfort \cite{chaurand2013cyclists,abadi2018bicyclist}. For instance, \cite{chaurand2013cyclists} studied the perceived risk of bicyclists and drivers in certain interactions. Results suggest that perceived risk is higher for drivers compared to bicyclists. Additionally, the perceived risk of bicyclists is higher when interacting with a car than with another bike. Another study investigated the perceived level of comfort by bicyclists near urban truck loading zones in varying conditions of truck traffic, bicycle lane marking type, and traffic signs \cite{abadi2018bicyclist}. Results indicate that the existence of trucks in the traffic is a significant factor in reducing bicyclists' perceived comfort. Additionally, the study finds that women are generally more affected by the truck traffic than men. While these types of studies add significant value to our understanding of the effect of contextual settings on bicyclists' perceived risk and comfort, they often are riddled with issues such as limitations on external validity. For instance, these studies may suffer from hypothetical bias, in which participants' responses to surveys may not reflect their real response in a naturalistic situation \cite{fitch2018relationship}. For instance, \cite{fitch2018relationship} reports that imagined ratings of comfort while biking may have a negative bias as high as 15\% difference in comfort and safety when compared to real-world situations. Additionally, surveys and subjective measures generally cannot be used to understand the temporal dimension of the effect of certain contextual elements on bicyclists. For instance, in the case of perceived comfort, it is not possible to understand the exact moment in which a bicyclist felt discomfort, or to what level the discomfort varies among different people and at different locations. On the other hand, physiological measures can be used as surrogate metrics to understand the time span of contextual elements' effects on the participants.   

\subsection{Observational Study} \label{sec:observational}
Observational studies can minimize the risk of hypothetical bias from stated preference surveys and provide a real-world assessment of bicyclists' responses in specific locations. For example, an observational study conducted in Boston, United States, investigated distracted cycling behavior and reported the prevalence of two types of distractions: auditory (ear buds/phones in or on ears), and visual/tactile (electronic device or other object in hand). Almost one-third of all bicyclists exhibited distracted behavior at four high traffic intersections during peak commuting hours. The highest proportion of distracted bicyclists was observed during the midday commute (between 13:30-15:00) \cite{wolfe2016distracted}. Another observational study with 2187 cyclists in Germany shows 22.7\% bicyclists are engaged in a secondary task such as wearing headphones or earphones (13.1\%) or interacting with other cyclists (7.0\%)  \cite{huemer2019secondary}.  

In observational studies, the collected data relies only on the behavioral responses that the observers can visually discern, without having the ability to manipulate different factors, such as traffic density or noise levels \cite{daniels2008effects}. In recent years, the utilization of cameras has greatly increased the popularity of video-based observational studies.  For instance, an observational study in China recorded 112 hours of video footage with 13,407 bicyclists riding shared bikes and 2061 riding personal bikes. Not wearing a helmet, violating traffic lights, riding in the opposite direction of traffic, not holding the handlebar with both hands, and riding in a non-bicycle lane are identified as top unsafe behaviors \cite{gao2020unsafe}. Overall, observational studies can only evaluate bicyclists' behaviors in existing environments, and are unable to collect bicyclist's psycho-physiological responses. Psycho-physiological measures of bicyclists may be helpful for understanding the reason behind distraction, the length of the distraction, and to what level the stimulus affects the rider's decision-making.  

\subsection{Naturalistic Study} \label{sec:real_world}
The development of mobile sensing technologies has enabled mobility data collection from a variety of modalities. In this line of research, a number of naturalistic studies have investigated bicyclists' behaviors and physiological responses, such as HR, heart rate variability (HRV) \cite{doorley2015analysis}, and gaze \cite{rupi2019visual}. For example, the changes in ambient light levels can affect bicyclists' perception of the environment by changing the gaze reactions \cite{uttley2018eye}. These studies show preliminary evidence on how infrastructure design is associated with cycling stress, although variations are found between different cities and road types \cite{fitch2020psychological, teixeira2020does,guo_robartes_angulo_chen_heydarian_2021}. In \cite{fitch2020psychological}, through using a BodyGaurd II heart beat-to-beat interval measuring device, the HRV results from 20 female participants suggest that only local roads (with no dividing yellow line and low car speed and volumes) consistently provided less stress to the participants compared to collector (medium to high volume road with striping for dividing traffic) and arterial roads (high volume, multi-lane). One of the limitations identified in the study is that environments with protected or separate bike lanes are not included, so the results cannot indicate how those designs might compare to the as-built designs \cite{fitch2020psychological}. Lack of environmental control (e.g., traffic, weather, etc.) could be the main cause of the uncertainty, which undermines the interpretability of these results and suggests the need for further research. Additionally, only a limited number of human sensing devices can be applied in naturalistic settings, as many of these devices are intrusive. For example, the electroencephalogram (EEG) measurement devices is more capable for in-lab testing. This will have an effect on participant's behavior and safety, which may result in degraded data quality. Lastly, the potential risks of injuries and fatalities for participants triggers ethical concerns for naturalistic studies. For example, \cite{stelling2018study} conducted a study in real traffic examining glance behavior of teenage bicyclists when listening to music. The study was terminated after fourteen participants when the results indicated that a substantial percentage of participants cycling with music experienced a decrease in their visual performance \cite{stelling2018study}.

\subsection{Bicycle Simulator With IVE Study} \label{sec:simulators}
The advancements in VR and bicycle simulation over the past decade has led to a rapid increase in its application among researchers, designers, and engineers to evaluate human responses to alternative infrastructure designs. The combination of IVE and instrumented physical bicycle simulators provides a high level of immersion and flexibility in experimental designs. Furthermore, it enables user engagement and allows subjective analysis of participants to better understand their behavior and preferences to the changes in simulated environments \cite{nazemi2018studying}. For instance, \cite{xu2017exploring} was able to evaluate 30 participants' behaviors in an IVE, by designing a straight path with four sections of varying traffic conditions. Through this experiment, results suggest that the existence of a bike lane in low traffic conditions significantly improved cyclist lane-keeping performance \cite{xu2017exploring}. A more recent study compared cycling behaviors in an IVE between a keyboard controlled bicycle and an instrumented bicycle where participants could pedal. The results indicated that there is more variance in the instrumented bicycle experiments in different measurements, such as speed, head movement, acceleration, and braking behaviors \cite{bogacz2020comparison}. Validation studies were also performed to compare the bicycling behavior between IVE bicycle simulator and naturalistic studies \cite{o2017validation,guo_robartes_angulo_chen_heydarian_2021}. Although there is a limited number of validation studies, promising results are shown in the validity of cycling performance such as lane position and speed \cite{o2017validation}. However, most IVE related studies are limited to observing cycling behaviors and preferences without exploring bicyclists' psycho-physiological responses.  

\subsection{Measurement of Physiological Responses} \label{sec:psycho_measure}
Past studies have pointed out the utility of humans' physiological signals such as cardiovascular (e.g., HR), skin temperature, skin conductance, brain signals, gaze variability, and gaze entropy in understanding emotions, stress levels, anxiety, and cognitive load \cite{tavakoli2021driver,kim2018stress,lohani2019review}. As we are focusing on wearable and eye tracking devices for this study, the following sections provide additional background on the correlation between stress level, cognitive load, HR, and gaze patterns. 

\subsubsection{HR/HRV}
Studies in different areas have used HR measures to analyze different human states in response to changes within an environment. While in medical applications HR is generally retrieved through devices such as Electrocardiography
(ECG), wearable based devices (e.g., smartwatches) generally use photoplethysmogram (PPG) technology. ECG measures the electrical activity of the heart through the application of contact electrodes, while PPG records the blood volume in veins using infrared technology. The blood volume measurement is then used to estimate the HR (i.e., beats per minute), and HRV \cite{lohani2019review,tavakoli2021harmony,tavakoli2021leveraging}. HRV features are a set of signal properties that are calculated based on the beat-to-beat intervals in a person's HR, such as the root mean squared of the successive intervals (RMSSD) \cite{tavakoli2020personalized,kim2018stress}. Both HR and HRV metrics are used in the literature for understanding human's state. In general, studies have shown that an increase in stress level is associated with an increase in HR, and a decrease in RMSSD features \cite{tavakoli2021harmony,kim2018stress,napoli2018uncertainty}. More specifically, in bicycling studies, an association has been found between perceptions of risk and HR \cite{doorley2015analysis,fitch2020psychological}. For instance, a naturalistic study in Ireland showed that situations bicyclists perceive to be risky are likely to elicit higher HR responses. This study also found that busy roads and roundabouts without bike lanes were perceived as more dangerous and risky compared to roads where cyclists are separated from traffic \cite{doorley2015analysis}. 

\subsubsection{Eye Tracking} \label{eye_tracking_definition}
In addition to HR, eye gaze patterns have also been used in different studies to infer human state. Different features such as blinking rate, saccade and fixation duration, gaze variability in different directions, stationary gaze entropy (SGE), and gaze transition entropy (GTE) have been shown to be correlated to different states such as work load, stress level, and emotions \cite{shiferaw2019review}. A fixation in eye patterns refers to maintaining the eye gaze on a specific location \cite{purves2001types}. At each fixation, the gaze is approximately stationary. The transitions between fixations are called saccades where the point of fixation changes rapidly to a new fixation point. The variation and sequence of fixations and saccades were shown to be correlated with human states such as stress and work load \cite{may1990eye}. In addition to fixation and saccade, gaze variability is an additional feature that refers to the standard deviation in gaze angles in both vertical and horizontal directions. 

In general, two measures can be calculated for entropy. The first one is based on the definition of uncertainty associated with a choice \cite{shiferaw2019review}. With more randomness in a system, the entropy also increases. This is calculated through Shannon's equation \cite{shannon1948mathematical}. In the gaze analysis research, this first entropy is referred to as the stationary gaze entropy (SGE), which shows the overall predictability of fixation locations and can be a proxy for the gaze dispersion \cite{shannon1948mathematical}. For a set of fixation locations in a sequence of eye movements, if we assign fixation locations to spatial bins of $p_i$, we can calculate the SGE as:
\begin{equation} \label{equation:sge}
SGE = -\sum_{\textit{i=1}}^{n} p_{i} \log_{2}p_{i} 
\end{equation}

Different studies have used SGE for human state analysis. For instance, SGE was used for detecting task demand, complexity, experience, workload, drowsiness, and being under the influence of alcohol \cite{shiferaw2019review}. 

The second measure of gaze entropy is gaze transition entropy (GTE), which is the conditional entropy that takes into account the temporal dependency between different fixations. GTE is a measure of predictability of the next fixation location given the current fixation location. For a sequence of transitions between different spatial bins of $i$ and $j$, with a probability of $p_{ij}$ the GTE can be calculated as:
\begin{equation}  \label{equation:gte}
GTE = -\sum_{\textit{i=1}}^{n} p_{i} \sum_{\textit{j=1}}^{n} p_{ij} \log_{2}p_{ij} 
\end{equation}

Conceptually, for each specific combination of task demand and scene complexity, an optimal level of GTE exists \cite{shiferaw2019review}. The optimal GTE can be imagined as the result of the interaction between human internal state and the amount of information provided by the external context. Deviation from the optimal GTE can provide information about changes in human state. For example, increase in stress, anxiety, and frequency of emotional episodes are associated with an increased level of GTE (relative to the optimum). While a decrease in the level of GTE (relative to the optimum) can be due to usage of depressants such as alcohol \cite{shiferaw2019review}. 

In addition to gaze entropy, gaze data is usually analyzed based on the Area of Interests (AOI). People divert their attention away from the previous fixation to another, which reflects the changes in mental concentration through AOI. By mapping fixations with the AOI, it is possible to obtain a statistical description of key gaze parameters, including the fixation duration, fixation counts, and saccade counts. In transportation related studies, road center is a frequently used AOI \cite{wang2014sensitivity}. The percentage of fixations in the road center AOI is referred to as the percentage of road center (PRC) feature  \cite{guo2020interacting}. PRC has been shown to increase with elevated cognitive demand \cite{engstrom2005effects}. 

For bicycle related studies, eye tracking behaviors have been analyzed in several real-world experiments. In a naturalistic study from Germany, 20 participants cycled at five defined test locations while wearing a mobile eye tracking system. The outcome shows that spatially open locations are related with higher level of perceived risk, and more cycling experience and greater familiarity with a location may lead to a more foresighted and focused gaze behavior \cite{von2020gaze}. Another naturalistic study in Italy investigated bicyclists’ eye gaze behavior at signalized intersections. They collected this data through a mobile eye tracker from 16 participants in a 3-kilometer corridor. The results show that intersections that force bicyclists to merge with vehicle traffic yield notable differences in features of gaze behavior. For instance, when approaching intersections, the moment of first fixation on the traffic lights occurs earlier for the case of no bike lane as compared to the case with separate bike lane. Additionally, for inexperienced cyclists, intersections without a separate bike lane were associated with an increase in gaze variability and looking around \cite{rupi2019visual}. To our knowledge, no previous research has studied bicyclist's gaze behaviors in IVE with bicycle simulators.

As previous studies have shown that human physiology (e.g., HR) is correlated with different changes in human states such as stress, comfort, and cognitive load, which may be used to infer bicyclists' behavioral changes with respect to the environmental variations \cite{tavakoli2021harmony,roe2020urban,tavakoli2021leveraging}. While previous studies provided significant insights into the effect of the road environment on bicyclists' state and behaviors, most of these studies did not explore the variation in bicyclists' physiological measures in response to changes in roadway design. In this study, by integrating a physical bicycle simulator within an IVE, we collect multimodal human sensing data from 50 participants through three different bicycle infrastructure designs on the same road. Through linear mixed effect modeling, we provide evidence on the strong association between bicyclists' psycho-physiological features and different roadway designs. 


\section{Methodology}
\begin{figure} 
    \centering
    \includegraphics[width=\linewidth]{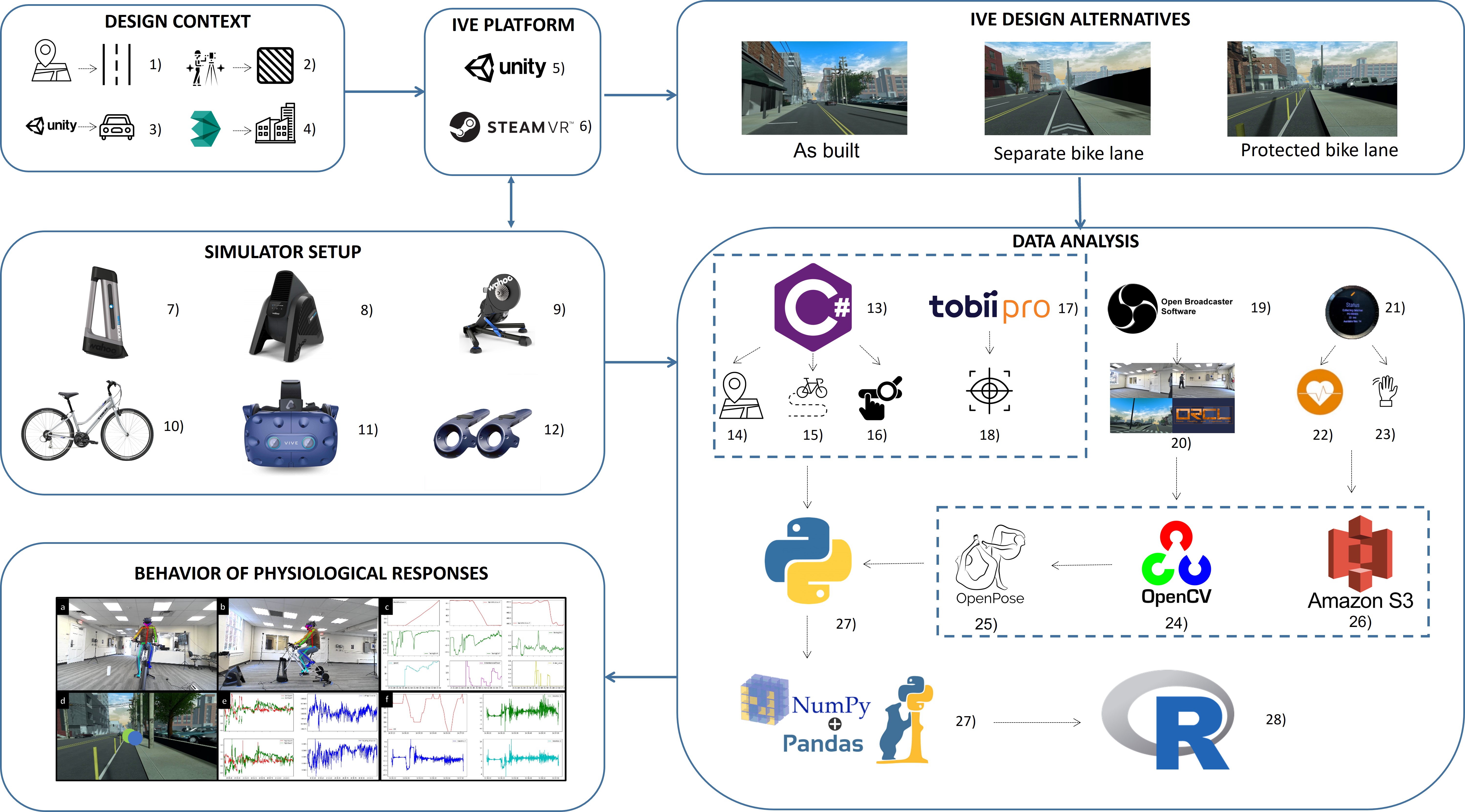}
    \caption{System architecture of data collection. Design context: 1) Road geometry information from Google map; 2) Road texture from real world measurement; 3) Vehicle modeling and traffic simulation in Unity asset store; 4) Buildings modeling from 3DMax. IVE Platform:  5) Unity: 3D gaming engine; 6) SteamVR: integrating hardware with Unity. Simulator Setup: 7) Wahoo Kickr Climber: simulates physical grade changes by adjusting bike incline; 8)  Wahoo Kickr headwind: headwind proportional to bike speed; 9) Wahoo Kickr Smart Trainer and ANT+: biking dynamics simulation; 10)  Trek Verve physical bike; 11) HTC VIVE Pro Eye: VR headset with eye tracking; 12) Controllers: steering and braking of the bike. Data Collection: 13) C\# scripts in Unity to record: 14) Position, 15) Cycling performance and 16) Interactions on controllers (touch, click or press); 17) TobiiPro Unity API collects: 18) Eye tracking data; 19) OBS studio: records room videos and VR videos simultaneously as shown in 20); 21) Android smartwatch collects: 22) Heart rate and 23) hand acceleration data. Data Preprocessing:  24) Opencv: video and image processing; 25) Openpose: pose data extraction from videos; 26) Amazon S3: smartwatch data on the cloud; 27) Python: numpy and pandas for data cleaning, management and analysis; 28) R: statistical modeling.}
    \label{fig:System_archetecture}
\end{figure}

The methodology section is divided into multiple subsections which describe the experiment design (section \ref{sec:experiment_design}), alternative scenarios (section \ref{sec:alternative}), bicycle Simulator setup (section \ref{sec:bicycle_simulator}), data collection (section \ref{sec:data_collection}), experiment procedure (section \ref{sec:experiment procedure}), participant recruitment (section \ref{sec:participant}), data preprocessing (section \ref{sec:preprocessing}), and statistical modeling (section \ref{sec:stat modeling}). Figure \ref{fig:System_archetecture} shows a general framework of this study.

\subsection{Experiment Design} \label{sec:experiment_design}
This research studies the effect of different roadway designs on bicyclists' physiological states. The independent variables are demographic information (i.e., age, gender, bicycling attitude, and VR experience), the subjective realism of the IVE, as well as the three categorical variables of different roadway designs in IVE with a bicycle simulator: (1) the as-built shared bike lane environment (sharrows), (2) separate bike lane, and (3) protected bike lane with pylons. The dependent variables are different measurements of cycling performance (i.e., speed, lane position) and physiological responses (i.e., eye tracking and HR features) from integrated or mobile sensors.

\subsection{Road Environment and Alternative Designs in IVE} \label{sec:alternative}
\begin{figure} 
    \centering
    \includegraphics[width=\linewidth]{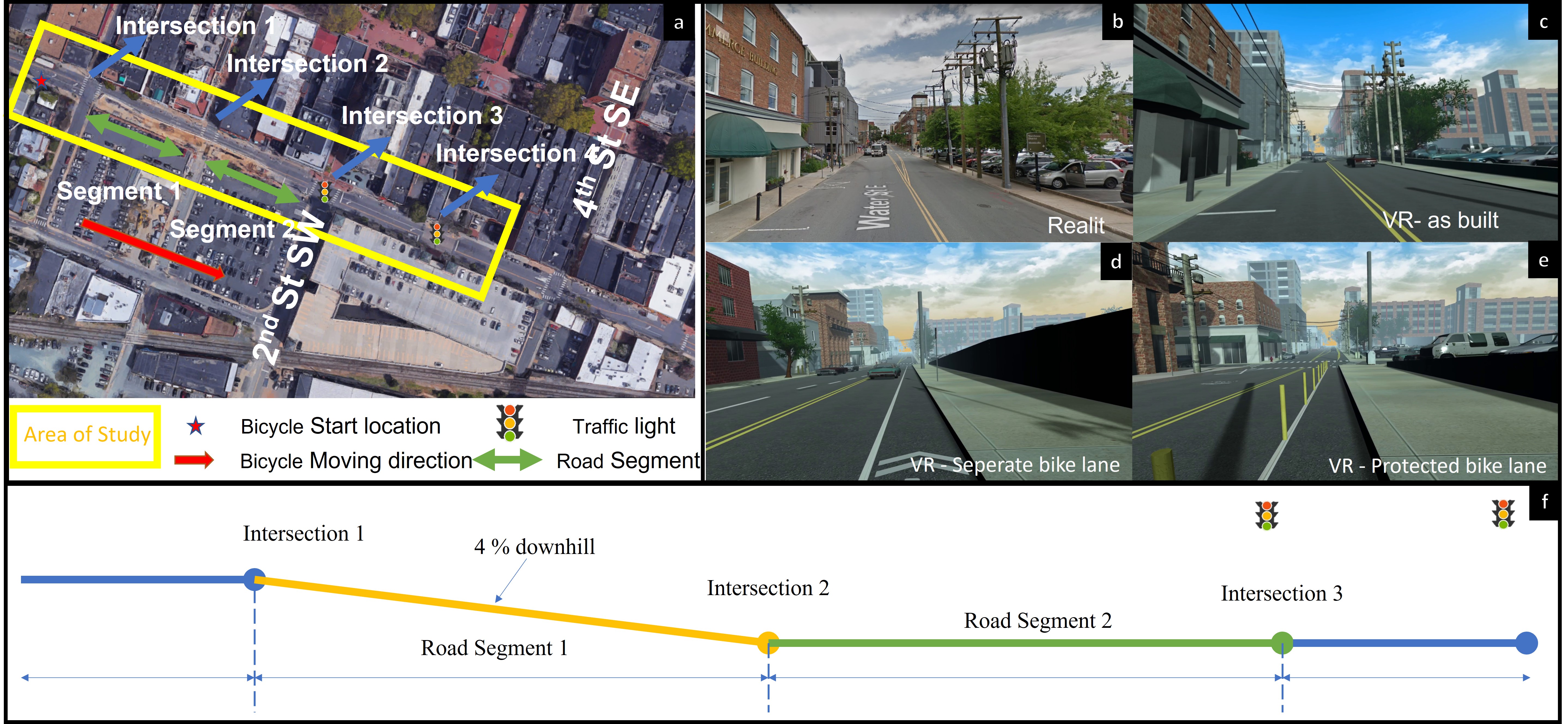}
    \caption{Real-world road design and alternative designs in IVE, (a) Real world location on the map; (b) Street view of the real world location; (c) Street view of as-built environment in IVE; (d) Street view of separate bike-lane environment in IVE; (e) Street view of protected bike lane environment in IVE; (f) the vertical road profile, road segment one is a 4\% downhill road}
    \label{fig:road}
\end{figure}

The IVE is developed based on a real-world location - the Water Street corridor in the city of Charlottesville, Virginia. This area has consistent volume of bicyclists and pedestrians and has been identified as a priority corridor by Virginia Department of Transportation for pedestrian crashes. The simulated road includes four city blocks, with a 4\% downhill grade in one of the segments (Figure \ref{fig:road} - (f)). The road has shared lane markings (sharrows) without separate bike lanes. At the third intersection, there is a traffic signal, and there is a parking lane in the westbound direction. Figure \ref{fig:road} shows a comparison between the real environment and the IVE created in Unity software and the location of the real-world environment. 

Based on technical drawings provided by the City of Charlottesville and in-field measurements  (\ref{fig:System_archetecture}-1), a one-to-one road environment is built in Unity (\ref{fig:System_archetecture}-5) with SteamVR platform (\ref{fig:System_archetecture}-6). The road textures are made from high resolution images of the real-world surfaces to make sure colors and surface details are representative of the real-world environment (Figure \ref{fig:System_archetecture}-2). The traffic volume is calculated from two weeks of real-world observations of the number of vehicles passing through the selected corridor. Based on the observations, four similar car models in Unity are generated randomly within the IVE to create traffic flows (Figure \ref{fig:System_archetecture}-3).
The buildings in the IVE are modeled individually in 3DMax software and then imported into the IVE (Figure \ref{fig:System_archetecture}-4). 
The bike lane width for both the separate bike lane and protected bike lane was designed to be 4ft (1.2m) wide, the minimum requirement based on Manual on Uniform Traffic Control Devices e(MUTCD) guidelines. Typically, bike lanes are to be between 4 to 6 feet (1.2 to 1.8m) wide. However, re-configuring the roadway markings for vehicular traffic and the inclusion of a bike lane only left enough space for one 5 foot (1.5m) bike lane on the roadway. Since this project aimed at having both a protected and separate bike lane, the standard bike lane width of 4 feet (1.2m) for both environments was deemed appropriate so that there would be no discrepancies between lane widths and user comfort during experimentation. The protected bike lane had an additional 1 foot (0.3m) wide buffer zone between the edge of the bike lane and the vehicular traffic with yellow pylons placed every 6 feet (1.8m).

\subsection{Bicycle Simulator Setup}\label{sec:bicycle_simulator}
The series of Wahoo indoor bicycling training equipment \cite{wahoowebsite2021} are used to build the physical simulator. The Kickr Climb can apply grade changes to the bike (Figure \ref{fig:System_archetecture}-7). The Kickr Headwind provides headwind in front of the bicyclist based on the speed (Figure \ref{fig:System_archetecture}-8). The Kickr smart trainer and the ANT+ are necessary to provide haptic feedback to the bicyclist, which is compatible with Unity and StreamVR for data communication (Figure \ref{fig:System_archetecture}-9). A physical Trek Verve bike (Fig.\ref{fig:System_archetecture}-10) is installed as the  body structure of the bicycle simulator to improve the realism. HTC Vive Pro Eye headsets (Figure \ref{fig:System_archetecture}-11) with the controllers (Figure \ref{fig:System_archetecture}-12) are used in the simulator. The spatial location of the controllers allows the system to detect turning movements, and the buttons on the controllers can be modified for braking action to control the bicycle speed. A more detailed description of the bicycle simulator setup is available in our previous study \cite{guo_robartes_angulo_chen_heydarian_2021, guo2021orclsim}. 

\subsection{Data collection} \label{sec:data_collection}
This section will introduce the data collection settings and procedures as demonstrated by the system framework in Figure \ref{fig:System_archetecture}.

\subsubsection{Cycling Performance}
As introduced above, Wahoo indoor bicycling training equipment are connected to Unity and SteamVR. With the scripts written in C\# programming language in Unity (Figure \ref{fig:System_archetecture}-13), it is possible to extract the position, speed and direction of any object in the IVE, including the headset, controllers, bicycle and other virtual objects such as vehicles at any given time (Figure \ref{fig:System_archetecture}-14,15). Meanwhile, the standby scripts also collect any input from the controllers such as the pulled trigger values (0 to 1), which represents the brake for the bike simulator (Figure \ref{fig:System_archetecture}-16). All the cycling performance data is collected per frame with system timestamp once the Unity starts any scenario. Once the experimental trial is completed, the text data will be saved locally in the computer for each participant. The Unity data is recorded at a frequency of 30 Hz.

\subsubsection{Eye Tracking}
The HTC VIVE Pro Eye has an integrated Tobii Pro eye tracker. With the Tobii Pro Unity SDK \cite{tobiiwebsite2021}, it can be utilized to collect eye tracking data for further data analysis (Figure \ref{fig:System_archetecture}-17). We have created documents and sample code about how to set up the Tobii Pro SDK in Unity \cite{xiangwebsite2021}.
The output of Tobii Pro raw data is the 3D gaze direction, gaze origin and pupil diameter (Figure \ref{fig:System_archetecture}-18). The frequency of eye tracking data is 120Hz. 

\subsubsection{Video Recording}
 As can be seen from bottom left of Figure \ref{fig:System_archetecture}(a,b), the two cameras are placed in positions where they can capture different angles of the bicyclist. The bicyclist's point of view in the IVE is monitored throughout the experiment. We use Open Broadcaster Software (OBS) Studio (\cite{obsstudio2021}, Figure \ref{fig:System_archetecture}-19) to integrate all video recordings simultaneously with a fixed frequency of 30 Hz and resolution of 1080p (1920×1080) for each video source (Figure \ref{fig:System_archetecture}-20). Video information for each video source (i.e., creation date, duration, height and width) can be extracted using windows file system information, which will be utilized for time synchronization. 

\subsubsection{Heart Rate Measurement}
Our system uses two Android smartwatches (one for each wrist) that are equipped with the “SWEAR” app \cite{boukhechba2020swear} for collecting long-term data from smartwatches (Figure \ref{fig:System_archetecture}-21). The SWEAR app records HR (1 Hz) and hand acceleration (100 Hz) (Figure \ref{fig:System_archetecture}-22,23). The data collection is turned on before the experiment by the researcher and then no further action is required. Both watches are connected to a smartphone via Bluetooth, the smartphone and computer are on the same wifi network, and the time of all devices are synchronized with the online server before the experiment. All data from the smartwatches will be stored on the local device and then uploaded to Amazon S3 cloud storage (Figure \ref{fig:System_archetecture}-26).

\subsubsection{Surveys}
A pre-experiment survey in the study aims to collect participants' demographic information (age, gender), their prior VR experience and what types of bicyclists they are (biking attitude). After the experiment, a post-experiment survey is followed to collect their experience in the IVE and their subjective safety preferences for different scenarios.

\subsection{Experiment Procedure} \label{sec:experiment procedure}
\begin{figure} 
    \centering
    \includegraphics[width=\linewidth]{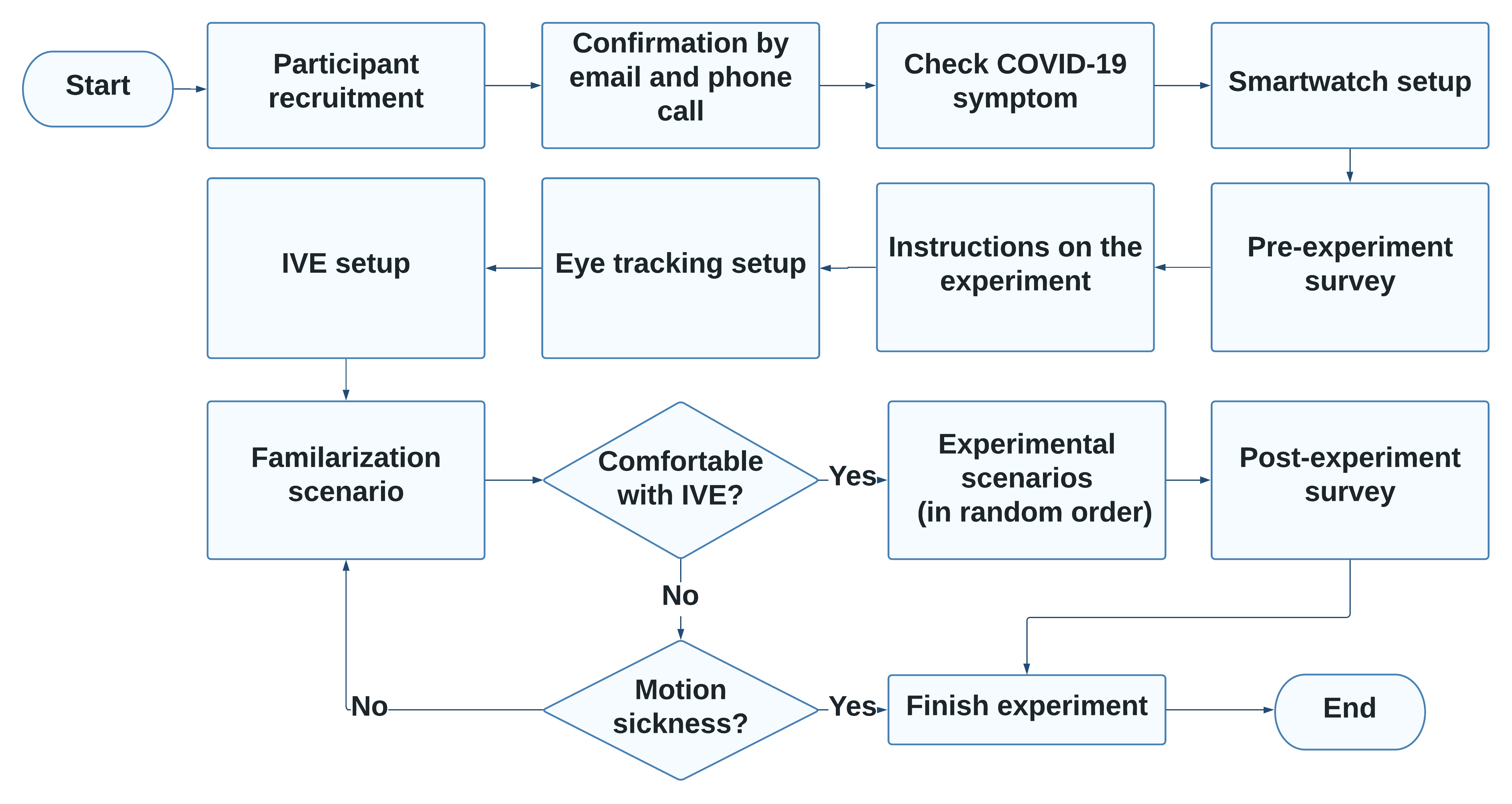}
    \caption{Experiment procedure}
    \label{fig:experiment_procedure}
\end{figure}

Figure \ref{fig:experiment_procedure} shows the experiment procedure. Once a prospective participant signed up, a researcher contacted the participant both via email and phone call a day before the experiment to confirm their reservation and health condition (due to the COVID-19 requirement). Upon arrival, each participant is asked to sign the consent form approved by the IRB office and put on two smartwatches on both wrists, before completing the pre-experiment survey. After finishing the pre-experiment survey, instructions are given on how to use the VR headset, controllers, and the bike simulator. After the bike is adjusted to a comfortable position, the participant is mounts the bike and puts on the headset. Next, the participant is guided through the eye tracker calibration. After the IVE system setup, the participant is placed into a familiarization scenario (without any vehicle traffic) to become accustomed to interacting with the IVE. In this environment, the participant can practice pedaling, steering, and braking, and the practice procedure can be repeated until the participant feels comfortable. If the participant feels any motion sickness, they may stop the experiment at any point and still receive compensation for participation. 

Once the participant is comfortable in the training environment, they experience the three design scenarios in random order, where each experiment trial lasts about two minutes, with a two minute break between each scenario. Once the participant has completed all three scenarios, they are asked to complete the post-experiment survey. On average, each participant spends 30 minutes completing the experimental procedure.

\subsection{Participants} \label{sec:participant}
\begin{figure} 
    \centering
    \includegraphics[width=\linewidth]{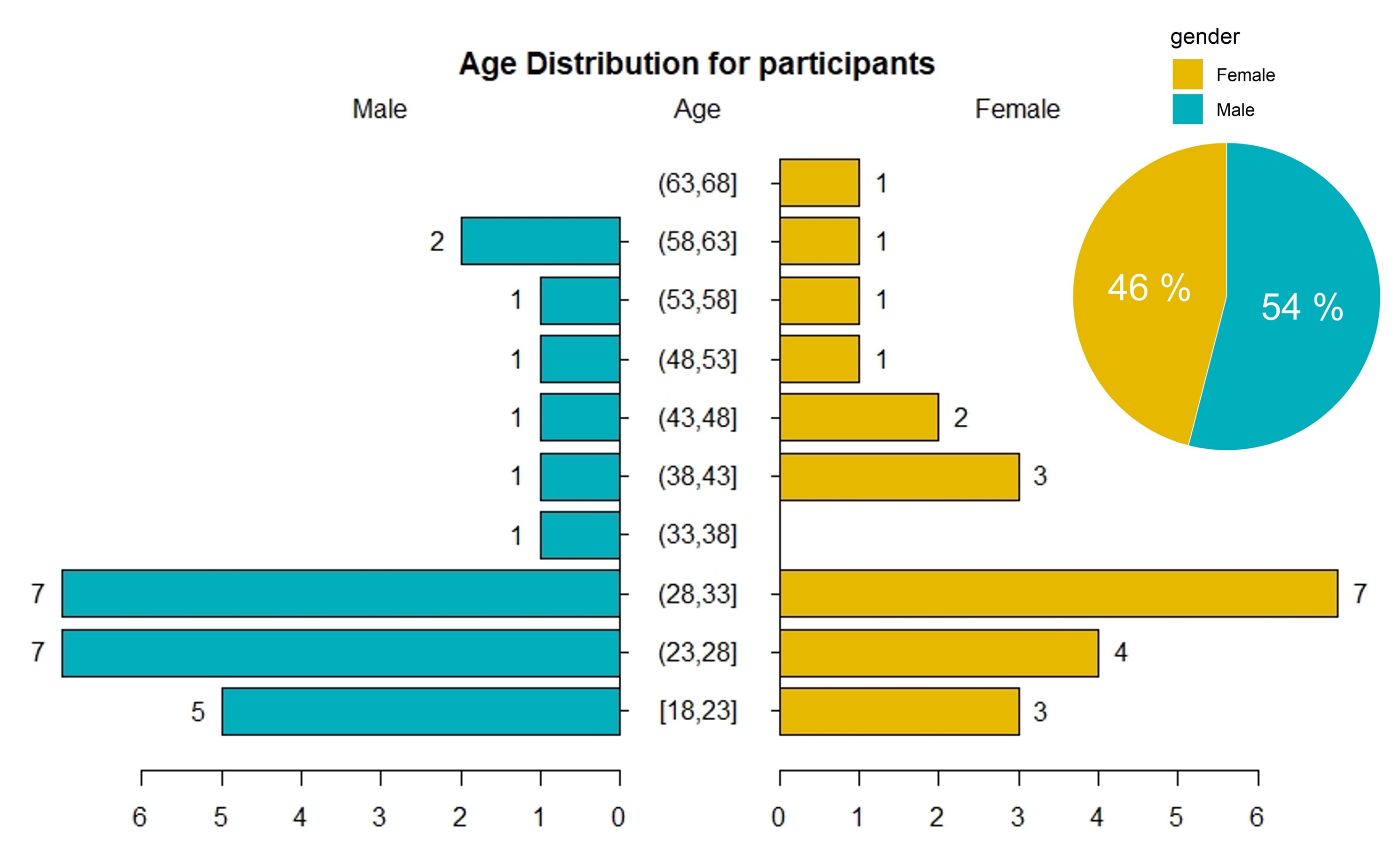}
    \caption{Age and gender distribution of participants based on the demographic data. Overall 23(46\%) are females and 27(54\%) are males. One male participant didn't reveal his age information}
    \label{fig:age_gender}
\end{figure}

51 participants were recruited for the experiment. Most of the participants are local bicyclists, university students and faculty members who are familiar with the study corridor. All participants are 18 or older and without color blindness. During the study, one participant could not finish the experiment due to motion sickness. For the remaining 50 participants (23 female and 27 male), the mean age is 34.1 with a standard deviation of 12.9 (1 participant did not reveal his/her age information); the age distribution is shown in Figure \ref{fig:age_gender}.

\subsection{Data Preprocessing} \label{sec:preprocessing}
This section introduces the necessary data preprocessing steps to extract the valid information, including time synchronization and feature extraction for the modeling section. 

Data extracted from different sensors have different frequencies. This required the resampling of each sensor separately: the cycling performance and video data is resampled at 30 Hz, the eye tracking data is resampled at 120 Hz, and the HR is resampled at 1 Hz. All data are trimmed to only include the data from the moment when the participants start pedaling until they pass the third intersection \ref{fig:road} in the IVE. The next step is combining the raw gaze direction from eye tracking data with the point of view videos, which helps transform the 3D gaze direction into 2D videos. This process allows us to visualize what the participants are looking at in the IVE. As shown in the lower left of Figure \ref{fig:System_archetecture}(d), the green and blue dots represent the left and right eye gaze points, respectively. The scripts for eye tracking data preprocessing are posted online \cite{xiangwebsite2021}.

\subsubsection{Fixations and Percentage of Road Center (PRC)}
Fixations are the most common feature of eye tracking to make inferences about cognitive processes or states. Fixations are the moments when eyes stop scanning about the scene and hold the central foveal vision in certain places to look for detailed information of the target object. We adapt our program from \textit{pygaze}, an open-source toolbox for eye tracking \cite{dalmaijer2014pygaze}, with 25 ms minimum duration and 100 pixel maximum dispersion thresholds to extract the fixation information from the 2D videos with gaze information. The mean fixation length is defined as the average length of all fixation events for any time interval of interest. From the distribution of fixations, the most frequent bins of fixations are considered as the road center, any fixations with a spatial distance less than 12$^\circ$ (as suggested by literature) will be taken as the road center fixations \cite{wang2014sensitivity}. To convert the 12$^\circ$ to actual pixel distance in the 2D gaze vector, the following equation is used (\ref{equ:distance}):

\begin{equation} \label{equ:distance}
D = \frac{tan(r)*w/2}{tan(FOV/2)}
\end{equation}

Where $r$ is the radius in angle of the road center, for which we choose 12$^\circ$, $w$ is the width of the field of view video, which is 1920 pixel. $FOV$ is the field of view angle of the headset, which is 110$^\circ$ for the HTC VIVE PRO eye. Therefore, the $D$ in our study is 142.88 pixel, which means any fixations with a euclidean distance less than $D$ will be considered as road center fixations. The PRC is defined as the percentage of fixation length within the road center among all fixations for any time interval of interest.

\subsubsection{Gaze Entropy}
As discussed in section \ref{eye_tracking_definition}, there are two types of gaze entropy measures: stationary gaze entropy (SGE) and gaze transition entropy (GTE). SGE provides a measure of overall predictability for fixation locations, which indicates the level of gaze dispersion during a given viewing period \cite{shiferaw2019review}. The SGE is calculated using equation (\ref{equation:sge}):
 
\begin{equation}
H(x) = -\sum_{i=1}^{n}(p_i)log_2(p_i) \tag{\ref{equation:sge}}
\end{equation}

$H(x)$ is the value of SGE for a sequence of data $x$ with length $n$, $i$ is the index for each individual state, $p_i$ is the proportion of each state within $x$. To calculate the SGE, the visual field is divided into spatial bins of discrete state spaces to generate probability distributions. Specifically, the coordinates are divided into spatial bins of 100 × 100 pixel. To get the trend of SGE, it is calculated in a rolling window of one second (120 data points in gaze raw data).

Gaze transition entropy (GTE) is retrieved by applying the conditional entropy equation to first order Markov transitions of fixations with the followed equation:
\begin{equation}
H_{c}(x) = -\sum_{i=1}^{n}(p_i) \sum_{i=1}^{n}p(i,j) log_2 p(i,j) \tag{\ref{equation:gte}}
\end{equation}

Here $H_{c}(x)$ is the value of GTE, and $p(i, j)$ is the probability of transitioning from state i to state j. The other variables have the same definitions as in the SGE equation (\ref{equation:sge}). 

\subsubsection{Change Points in HR}
In order to understand the abrupt changes in bicyclists' HR, we perform change point detection through Bayesian Change Point (BCP) analysis. As previously noted, an increase in HR can be associated with an increase in stress level, anxiety, and possibly negative emotions. Using a change point detector we can detect the moments in the time series of a bicyclist's HR that can be associated to increased stress levels in the biking scenario. This has also been explored in previous studies using BCP for HR analysis \cite{tavakoli2021multimodal,tavakoli2021harmony,guo2021orclsim}. In order to perform BCP, we use the Bayesian Change point model in Barry and Hartigan's book \cite{barry1993bayesian}. This model assumes a constant mean for the data and attempts to detect the point in which the mean changes. We perform this process in R \cite{team2013r} using the package \textit{bcp} \cite{erdman2007bcp}. 

\subsection{Statistical Modeling} \label{sec:stat modeling}
a Linear Mixed Effects Model (LMM) was chosen to model the different response variables across participants. LMM facilitates the analysis as it has the ability to (1) characterize group and individual behavior
patterns in a formal way, (2) acknowledge
both group and individual differences, and (3) incorporate
additional covariates \cite{krueger2004comparison}. 

LMM is centered around the idea of variability across participants \cite{brown2021introduction,fox2002linear}. Similar to simple linear regression, a LMM estimates the fixed factor effect (e.g., the effect of different scenarios on participant's heart rate and gaze measures). A LMM also estimates random effects, which arises from individual differences across participants. For instance, the effect of having a bike lane on participant's HR varies by participant (e.g., one participant experiences 20 percent increase in HR while another participant experiences 10 percent). A LMM is defined as:

\begin{equation}\label{lmm}
    y = X\beta + bz +\epsilon
\end{equation}

In equation (\ref{lmm}), $y$ is the dependent variable (e.g. heart rate or gaze measures), $X$ is the matrix of predictor variables, $\beta$ is the vector of fixed-effect regression coefficients, $b$ is the matrix of random effects, $z$ is the coefficients for each random effect, and $\epsilon$ is the error term of unexplained residuals. Additionally, the elements of the $b$ and $\epsilon$ matrices are:
\begin{equation}
    b_{ij} \sim N(0,\psi_k^{2}),Cov(b_k,b_{k'})
\end{equation}

\begin{equation}
    \epsilon_{ij} \sim N(0,\sigma^{2}\lambda_{ijj}),Cov(\epsilon_{ij},\epsilon_{ij'})
\end{equation}

The LMM is applied using the lme4 package in R \cite{bates2007lme4}. 

To evaluate participants' physiological responses in different road environments, as shown in Table \ref{table:variable definition}, different behavioral and physiological responses are treated as dependent variables in each LMM model, including cycling performance (speed and lateral lane position), eye tracking metrics (SGE, GTE, PRC, and mean fixation length), and HR metrics (mean HR and number of HR change points). Independent variables include type of bicycle infrastructure (as-built, separate bike lane, and protected bike lane), age (older or younger than 30), attitude towards cycling (based on pre-survey response), prior VR experience, and participants' sense of realism for bike speed and braking in the IVE. Additionally, each participant is treated as a random effect in the model. If statistically significant effects are revealed in any LMM models for the scenario variable, post hoc contrasts will be performed for multiple comparisons using Fisher's Least Significant Difference (LSD) \cite{williams2010fisher}.
All statistical analyses were performed at a 95\% confidence level ($\alpha = 0.05$).

\begin{table}[h!]
\caption{Summary of independent and dependent variables in LMM models}
\label{table:variable definition}
\centering
 \begin{tabular}{c c c c } 
  \hline
 Variable Type & Variable Name  & Categories  & Data Source \\ 
 \hline
 & Scenario & 3 & IVE Design \\ 
            & Age & 2 & Pre-survey\\ 
    Independent     & Gender & 2 & Pre-survey \\
   (Categorical)    & Type of bicyclist & 4 & Pre-survey \\
      & VR experience & 4 &  Pre-survey\\
            & Realism of bike steering & 5 & Post-survey\\
            & Realism of bike speed & 5 & Post-survey\\
  \hline
  & Cycling speed & km/h & Unity \\ 
           & Lateral lane position & m & Unity \\ 
           & Horizontal gaze variability & pixel & Eye Tracking \\ 
    Dependent       & Percentage of road center gaze & percentage & Eye Tracking \\ 
   (Continuous)    & Mean fixation length & second & Eye Tracking \\ 
     & Stationary gaze entropy & bits & Preprocessing \\ 
           & Gaze transition entropy & bits & Preprocessing \\ 
           & HR & bpm & Smartwatch \\ 
           & Number of HR change points & count & Preprocessing \\ 
 \hline
 \end{tabular}
\end{table}

\section{Results}
This section reports the results of the experiment. The following subsections describe the summary statistics (from the pre- and post-experiment surveys), the bicyclists’ physical behavior (cycling speed and lateral lane position), and physiological responses (eye tracking and HR) in different roadway designs. 

\subsection{Survey Response}
\subsubsection{Pre-experiment Survey}
\begin{figure} 
    \centering
    \includegraphics[width=\linewidth]{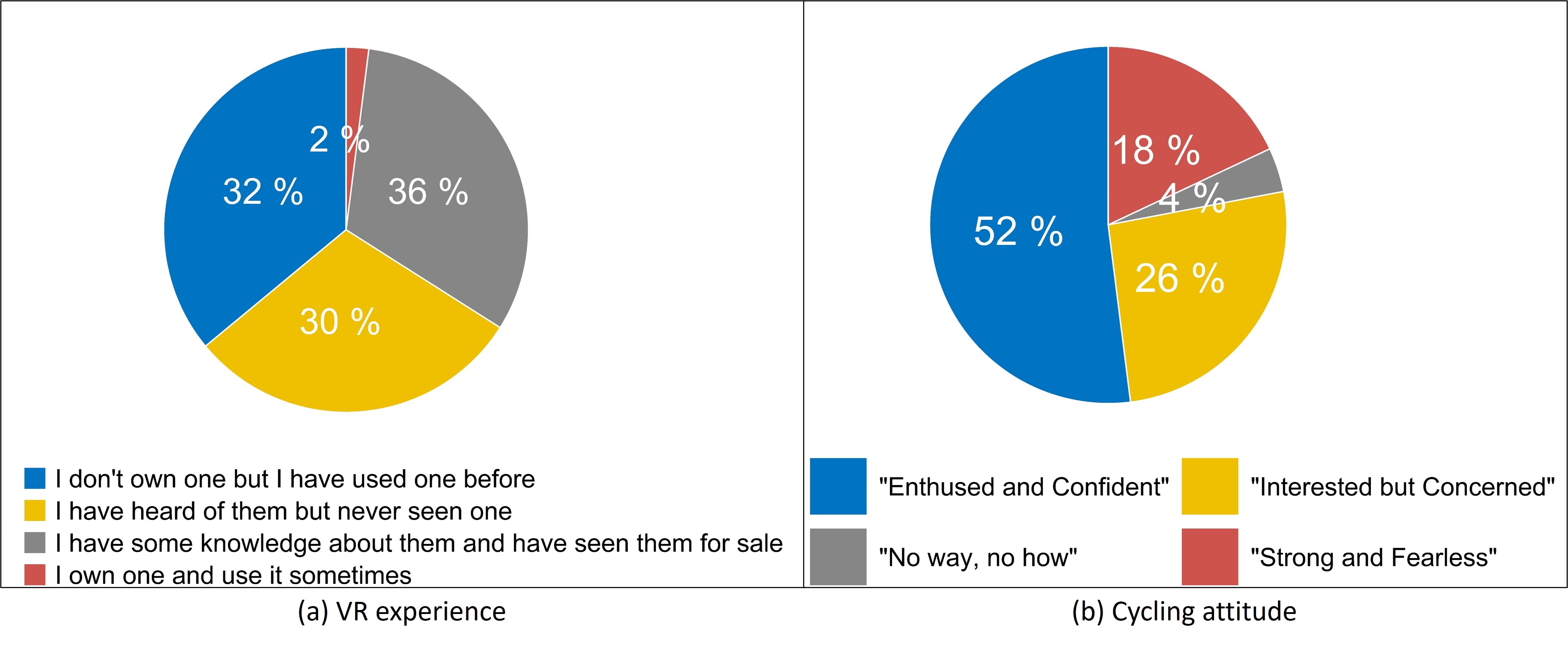}
    \caption{Summary of some survey responses. (a) Prior knowledge on VR, (b) Type of bicyclist}
    \label{fig:survey}
\end{figure}

All participants indicated that they have some level of prior knowledge of VR, although only one participant owns VR equipment and uses it regularly, as shown in Figure \ref{fig:survey}-a. The majority of the participants have a positive attitude towards cycling, as shown in Figure \ref{fig:survey}-b, with only two participants expressing hesitancy of cycling under any condition.



\subsubsection{Post-experiment Survey}
\begin{figure} 
    \centering
    \includegraphics[width=\linewidth]{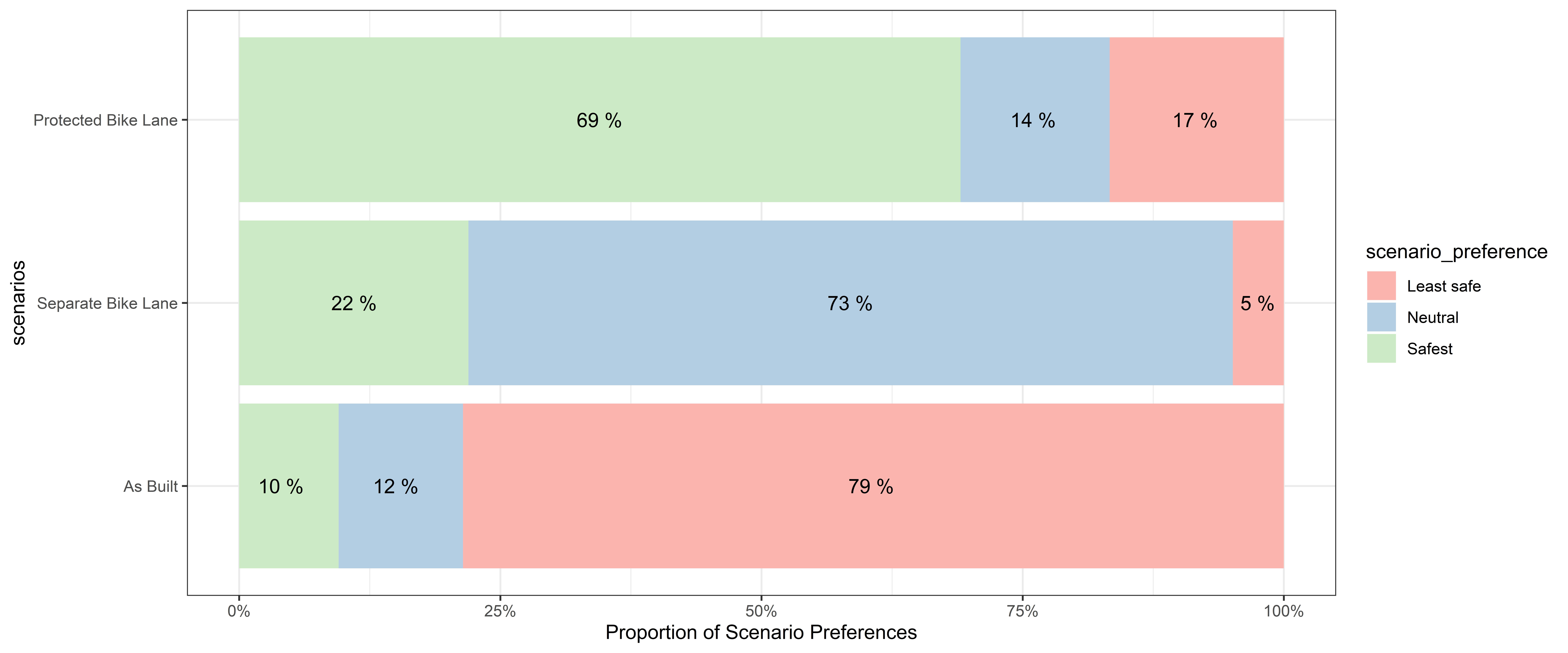}
    \caption{Scenario preference across participants based on the post-experiment survey. Note that 94\% of participants choose a 4 or 5 on the 5-point Likert scale (mean=4.42), with 5 indicating a full immersion }
    \label{fig:preferences}
\end{figure}

In the post-experiment survey, the majority of participants indicated that the virtual environment was immersive, with 94\% of participants choosing a 4 or 5 on the 5-point Likert scale (mean=4.42), with 4 and 5 indicating "immersed" and "very immersed", respectively. Most participants also found that the virtual environment was to scale (94\% chose 4 or 5, mean=4.54). The participant's feelings of speed and steering realism were both above average with 50\% indicating a 4 or 5 level of realism (mean = 3.56), and 54\% indicating a 4 or 5 for the steering realism (mean = 3.60). Fig.\ref{fig:preferences} shows the results of participants' scenario preferences. Participants indicated an overwhelming preference towards the protected bike lane (69\% rate it as the safest), followed by separate bike lane (22\% rate it as the safest), and the as-built scenario rated the least preferred safe environment (10\% rate it as the safest).

\subsection{Cycling Performance}
\begin{figure} 
    \centering
    \includegraphics[width=\linewidth]{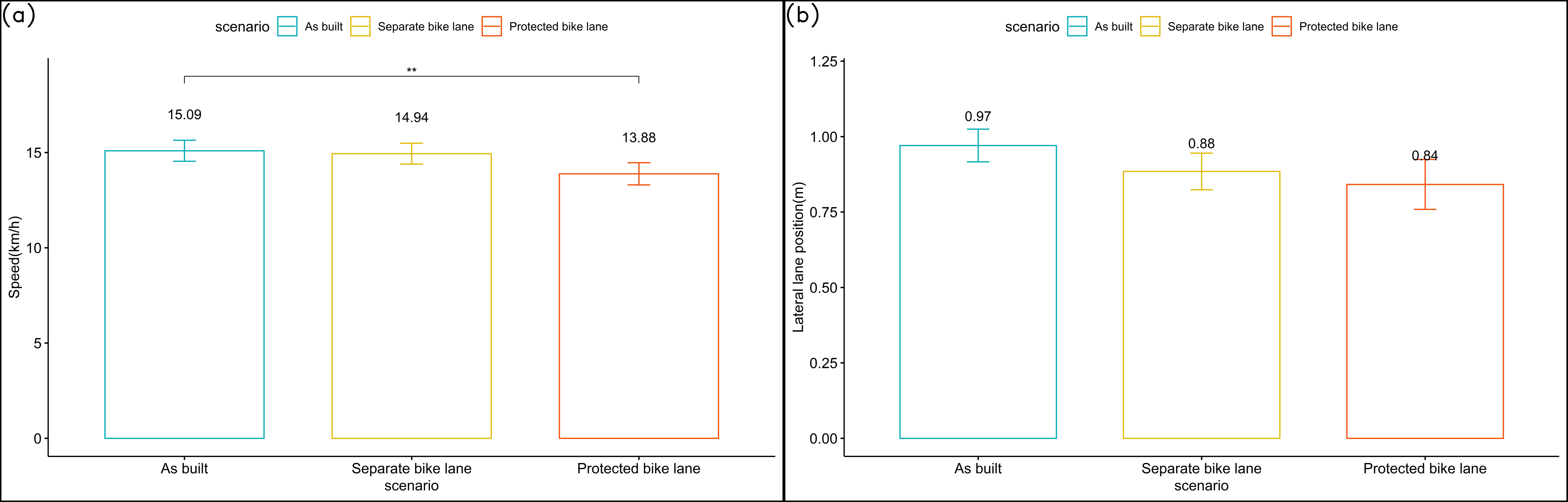}
    \caption{Cycling performance measured through speed (a) and lateral position (b) across different scenarios. Note that there is a trend for participants to move closer to the road curb to stay away from vehicles with the presence of separate bike lane or protected bike lane with pylons.  }
    \label{fig:cycling_performance}
\end{figure}

Two LMM are built individually for average speed and lateral lane position to estimate the relationship between the independent variables and participants' cycling performance.
For the mean speed LMM, there is a significant difference between the as-built and protected bike lane scenarios ($\beta = -1.209, SE = 0.383, p < 0.01$). Bicyclists' mean speed in the protected bike lane with pylons scenario (13.88 km/h) is significantly lower compared to the as-built scenario (15.09 km/h). No significant differences are found between the separate bike lane scenario (14.94 km/h) and the as-built scenario, as shown on Figure \ref{fig:cycling_performance} - a. Similarly, there is no significant difference in participants' speed between the bike lane scenario and protected bike lane with pylons. The random effect for the mean speed model is significant ($\beta =-1.209, SE = 0.383, p < 0.001$), suggesting  that  it  is  necessary  to treat the participant as a random factor in the model. This is also indicative of individual differences across participants in their cycling performance. 

For the mean lateral lane position, no significant differences are found across the three scenarios, although the difference between as-built and protected bike lane with pylons scenarios are marginally significant ($\beta =-0.129, SE = 0.070, p=0.068$). As shown in Figure \ref{fig:cycling_performance} - b, the average distance to the roadside curb for the three scenarios (as-built, separate bike lane, and protected bike lane with pylons) are 0.97 m, 0.88 m and 0.84 m respectively. The greater the average distance to the curb is, the smaller lateral distance between the bicycle and the vehicles. Therefore, there is a trend of participants moving closer to the curb to stay away from vehicles with the presence of separate bike lane or protected bike lane with pylons. The random effects for this model is significant ($\beta =0.971, SE = 0.067, p < 0.001$) as well.

\subsection{Eye Tracking}
\begin{figure} 
    \centering
    \includegraphics[width=\linewidth]{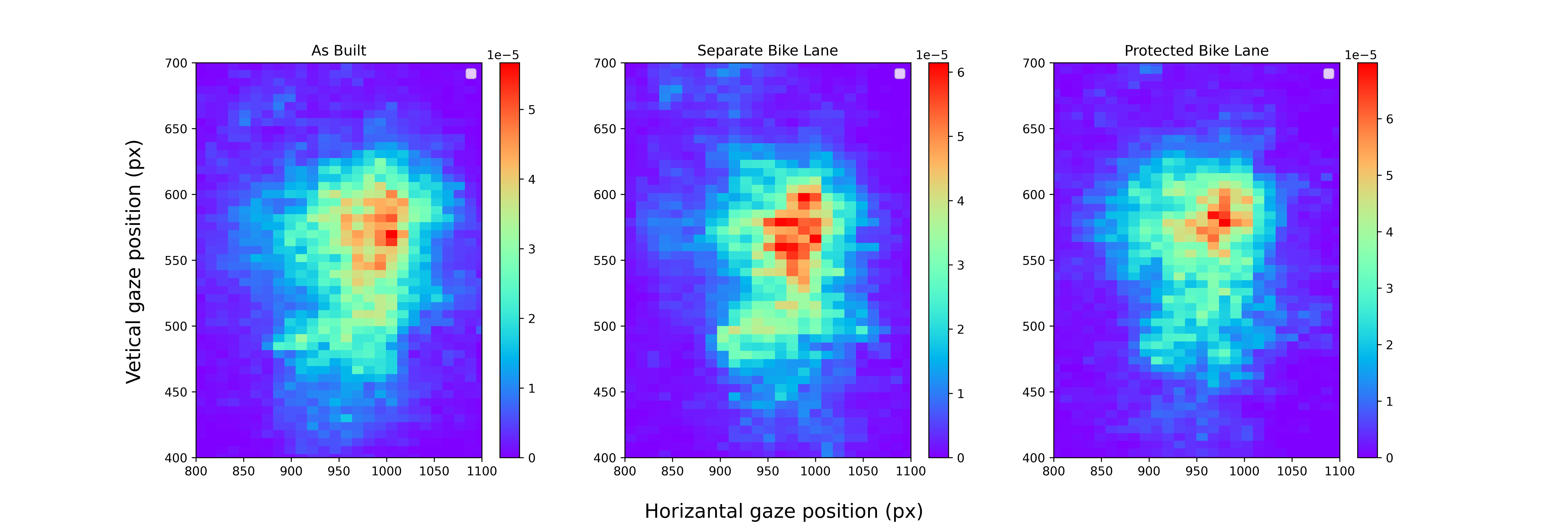}
    \caption{Gaze density heat map for different scenarios. Note that visual observations from the gaze heat map indicate that the as-built scenario has a more dispersed distribution than the other two scenarios}
    \label{fig:gaze_density}
\end{figure}

Five LMM are built individually for each eye tracking dependent variable (Table \ref{table:variable definition}) to estimate the relationship between the independent variables and participants' eye tracking metrics (horizontal gaze variability, PRC, mean fixation length, SGE, and GTE). We first plot the eye tracking heat map in the field of view to get an overview of the gaze distribution. As shown in Figure \ref{fig:gaze_density}, visual observations from the gaze heat map indicate that the as-built scenario has a more dispersed distribution than the other two scenarios. The separate bike lane scenario appears to have a higher concentration in the center of the gaze area, followed by the protected bike lane with pylons scenario. 

\subsubsection{Horizontal Gaze Variability}
\begin{figure} 
    \centering
    \includegraphics[width=\linewidth]{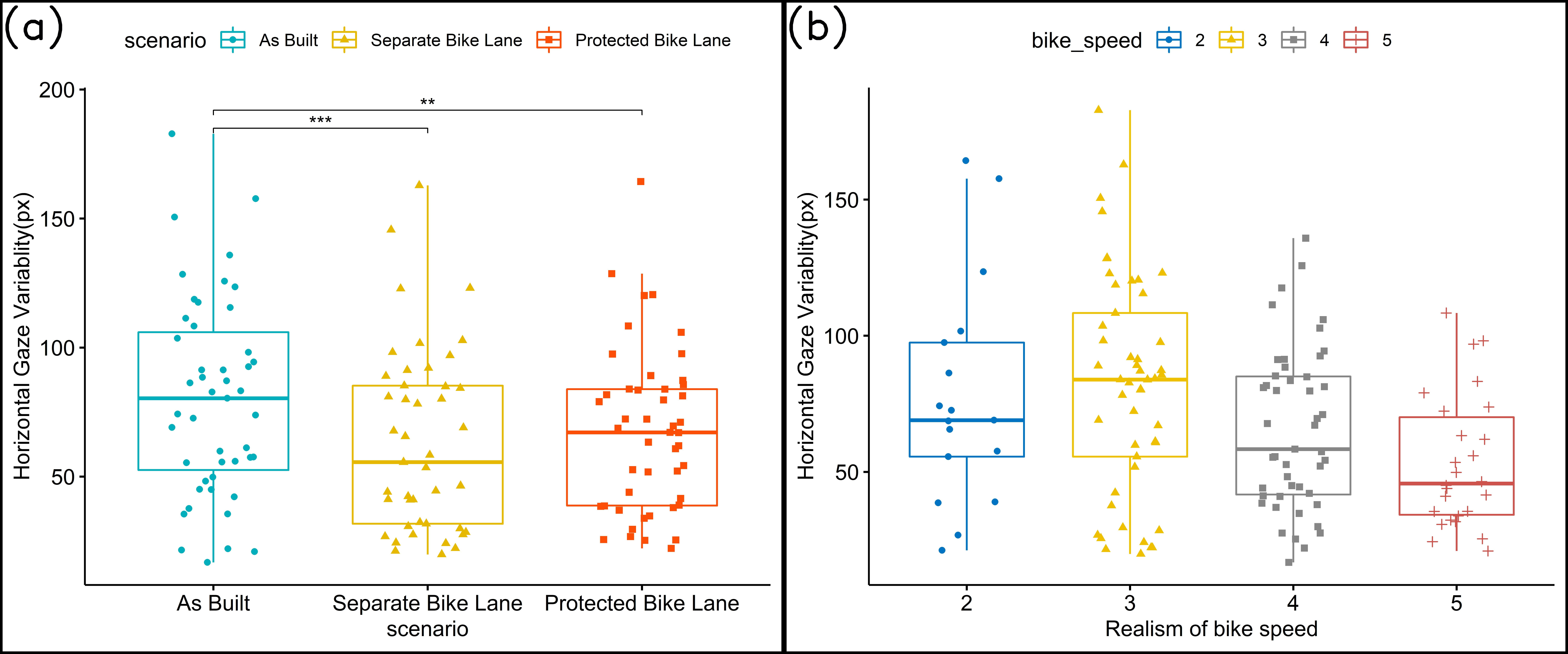}
    \caption{Horizontal gaze variability within different scenarios (a) as well as within different ratings of the realism of the bike speed (b). Note that a significant lower horizontal gaze variability is achieved both in the separate bike lane and protected bike lane. Additionally, the higher realism of bike speed the participants indicate, the lower horizontal gaze variability they show during the experiment}
    \label{fig:gaze_x_scenario}
\end{figure}

As illustrated above, a LMM is built for evaluating the relationship between participants' horizontal gaze variability and the independent variables. In Figure \ref{fig:gaze_x_scenario}, the result of the horizontal gaze variability model shows the random effects were significant ($\beta = 86.257, SE =32.990, p < 0.05$), which suggests that it was necessary to treat the participant as a random factor in the model. Both the separate bike lane and protected bike lane scenarios are statistically significant predictors for the horizontal gaze variability ($\beta = -16.349 , SE = 4.288, p< 0.001$ and $\beta = -12.645, SE = 4.278, p< 0.01$, respectively). As shown in Figure \ref{fig:gaze_x_scenario} - a, a significant lower horizontal gaze variability is observed both in the separate bike lane and protected bike lane, which indicates that participants are more focused directly ahead rather than laterally looking around the road environment. Another significant factor revealed by the model is the realism score of the bike speed from the post-experiment survey ($\beta = -13.991, SE = 6.287, p < 0.05$). Generally speaking, the higher realism of bike speed the participants indicate, the lower horizontal gaze variability they show during the experiment (Figure \ref{fig:gaze_x_scenario} - b, except for the small group who selected 2). No significant results are found in terms of the steering realism score. 

\subsubsection{Percentage of Road Center Fixation}
\begin{figure} 
    \centering
    \includegraphics[width=\linewidth]{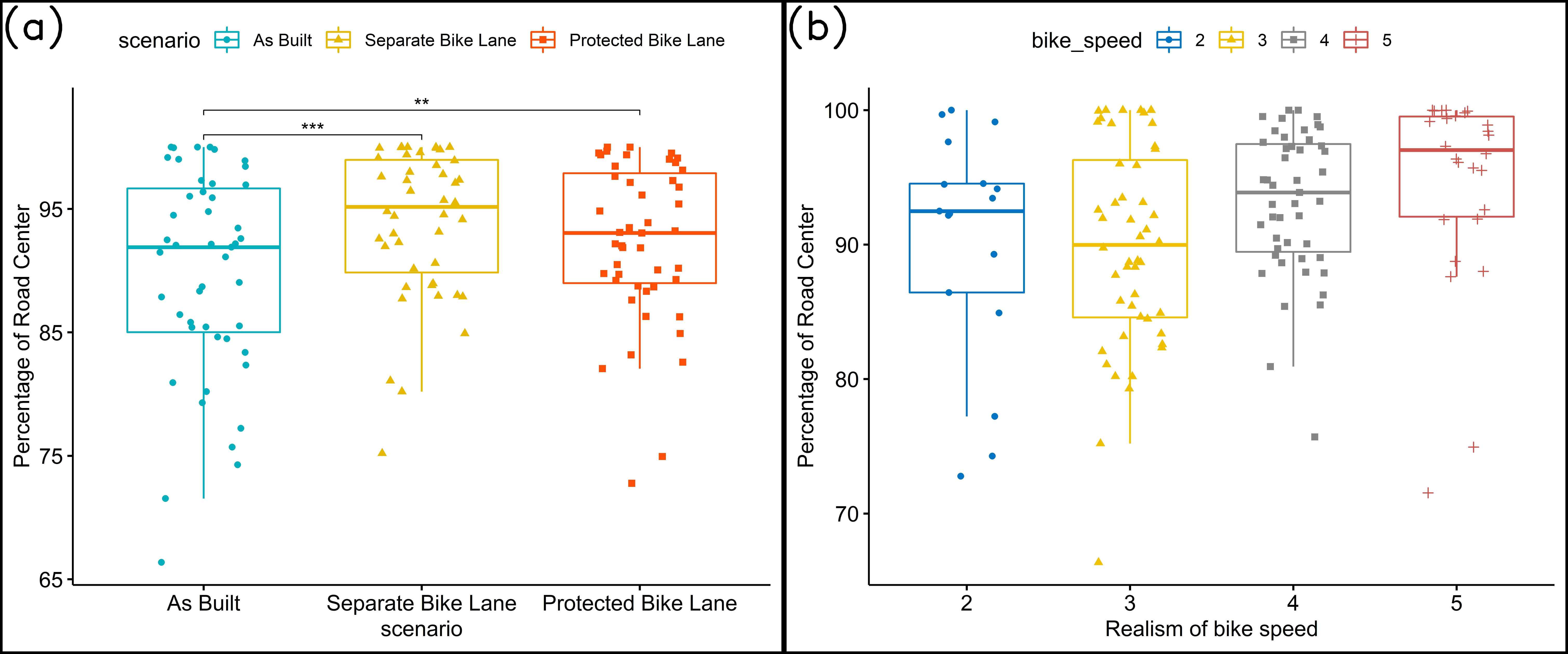}
    \caption{PRC and mean fixation length within different scenarios. Note that the percentage of road center fixation in separate bike lane is slightly higher than the protected bike lane, which aligns with the visual observation in gaze heat map }
    \label{fig:fixation_PRC}
\end{figure}

A similar LMM is built for the percentage of road center fixation. As shown in Figure \ref{fig:fixation_PRC}, a similar result is presented by the LMM for the horizontal gaze variability; the random effects are also significant ($\beta = 91.993, SE = 6.045, p < 0.001$). For the independent variables, both the separate bike lane ($\beta = 4.083, SE = 0.947, p< 0.001$) and protected bike lane ($\beta = 2.558, SE = 0.938, p< 0.01$) scenarios are statistically significant. The percentage of road center fixation in the separate bike lane is slightly higher than the protected bike lane, which aligns with visual observation of the gaze heat map (Figure \ref{fig:gaze_density}). This result indicates participants focus their gaze most on the road center in the separate bike lane. The realism score of the bike speed from the post-experiment survey is also significant ($\beta = 2.892, SE = 1.218, p < 0.05$). 

\subsubsection{Mean Fixation Duration}
\begin{figure} 
    \centering
    \includegraphics[width=\linewidth]{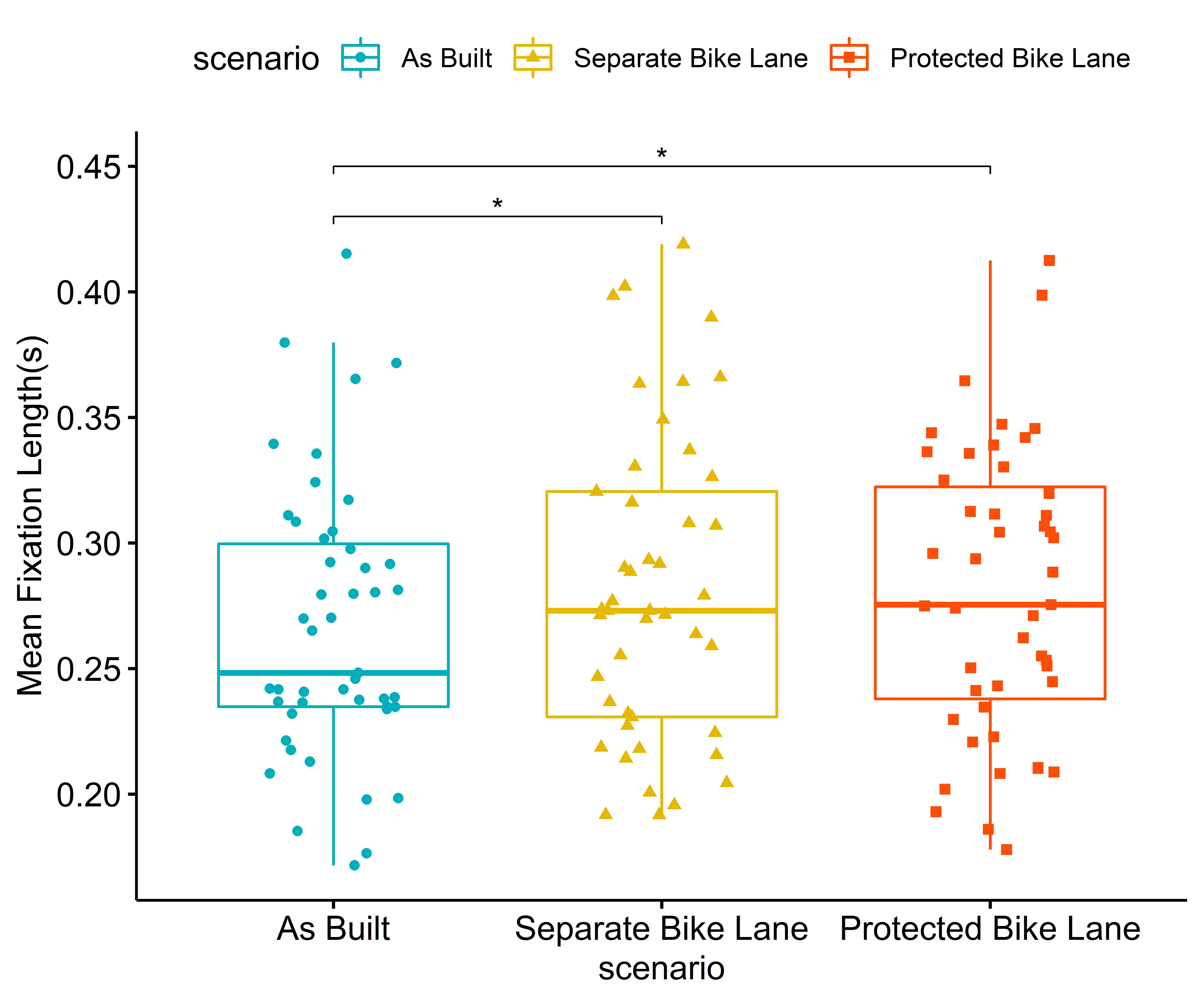}
    \caption{Mean fixation duration within different scenarios. A significantly higher fixation duration is observed both in the separate bike lane and protected bike lane scenarios compared to the as-built scenario.}
    \label{fig:meanfixation}
\end{figure}

The LMM model for the mean fixation duration shows the random effects are significant ($\beta = 0.242, SE = 0.051, p < 0.001$), and both the separate bike lane and protected bike lane scenarios are statistically significant predictors of mean fixation duration ($\beta = 0.015, SE = 0.007, p< 0.05$ and $\beta = 0.014, SE = 0.007, p< 0.05$, respectively). As shown in Figure \ref{fig:meanfixation}, a significantly higher fixation duration is observed both in the separate bike lane and protected bike lane scenarios compared to the as-built scenario. No significant results are found for other independent variables. 

\subsubsection{Gaze Entropy}
\begin{figure} 
    \centering
    \includegraphics[width=\linewidth]{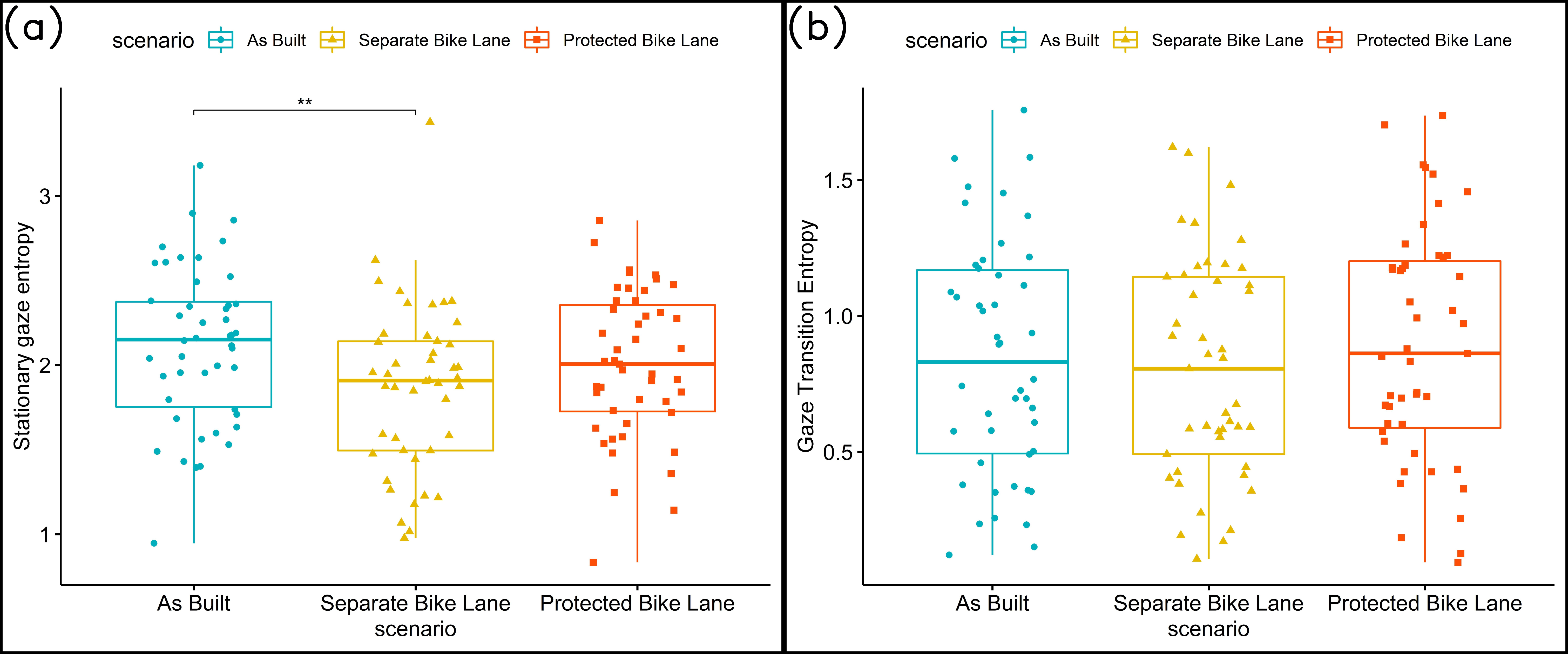}
    \caption{Stationary gaze entropy (a), and gaze transition entropy (b) within different scenarios. Note that the SGE in the separate bike lane environment is significantly lower than the as-built environment.}
    \label{fig:gaze_entropy}
\end{figure}

Two LMM are built for SGE and GTE. In both models, the random effects are significant ($\beta = 1.951, SE = 0.413, p< 0.001$ and $\beta = 0.791, SE = 0.306, p< 0.05$, respectively). Other than the random effects, only the separate bike lane in the SGE model is a significant predictor ($\beta = -0.224, SE = 0.079, p< 0.01$). As shown in Figure \ref{fig:gaze_entropy} - a, the SGE in the separate bike lane environment is significantly lower than the as-built environment. No significant results are observed in the GTE model.

\subsection{HR}
\subsubsection{Mean HR}
\begin{figure} 
    \centering
    \includegraphics[width=\linewidth]{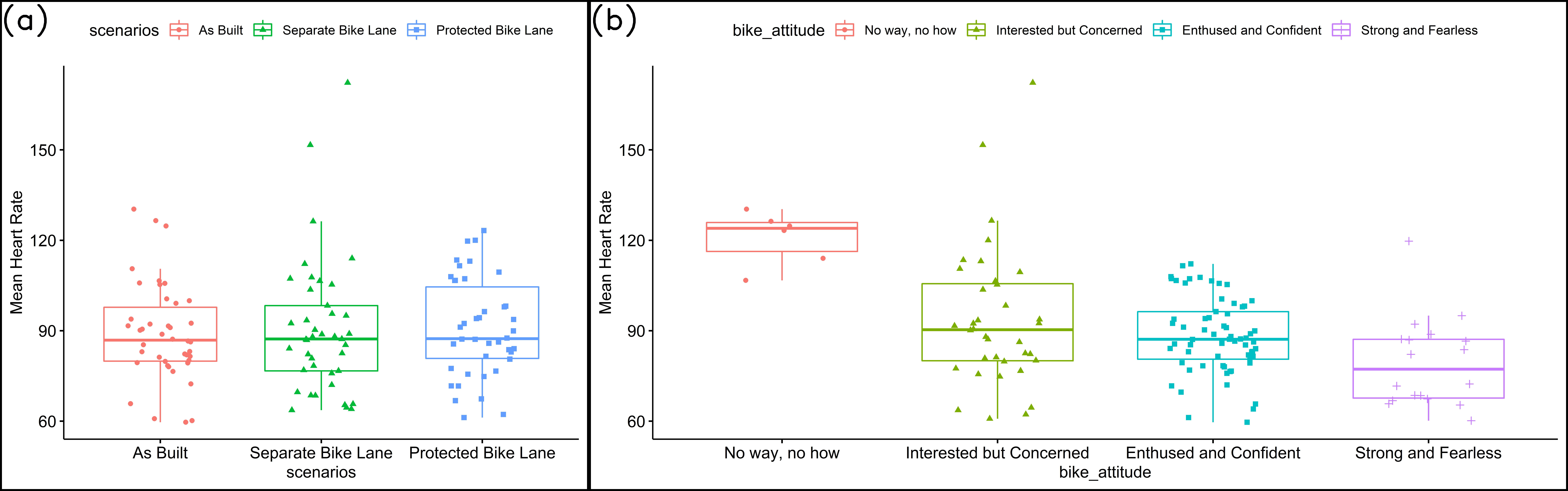}
    \caption{Mean HR within different scenarios (a) as well as attitude towards biking (b). Note that the more positive attitude participants have on biking, the lower mean HR they have during the experiment. }
    \label{fig:mean_HR}
\end{figure}

An LMM is built for mean HR during each experiment to compare the overall HR levels in different infrastructure designs. The random effects are significant ($\beta = 92.892, SE = 29.191, p< 0.01$). No significant results are found between different road designs (Figure \ref{fig:mean_HR} - a). A significant result is shown in type of bicyclist attitude based on the survey result ($\beta = -8.410, SE = 4.082, p<0.05$). As shown in Figure \ref{fig:mean_HR} - b, the more positive attitude participants have on biking, the lower mean HR they have during the experiment. Note that only two participants responded to the bicycling attitude question with 'No way, no how' so only six data points for this category are available.

\subsubsection{HR Change Point}
\begin{figure} 
    \centering
    \includegraphics[width=\linewidth]{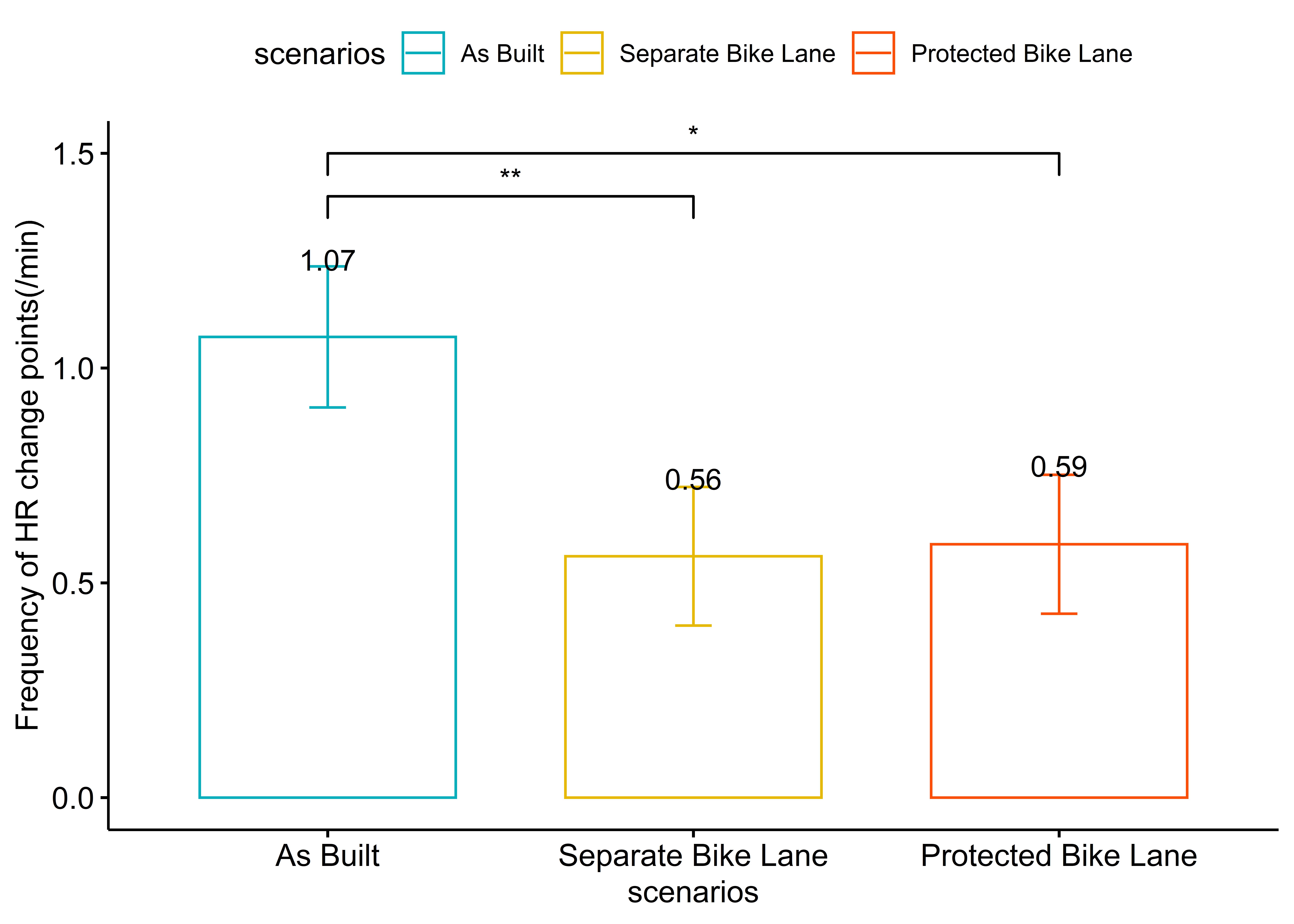}
    \caption{Frequency of HR change points within different scenarios. Note that both the separate bike lane and protected bike lane have significantly lower frequency of HR change point than the as-built scenario.}
    \label{fig:hr_change_point}
\end{figure}

\begin{figure} 
    \centering
    \includegraphics[width=\linewidth]{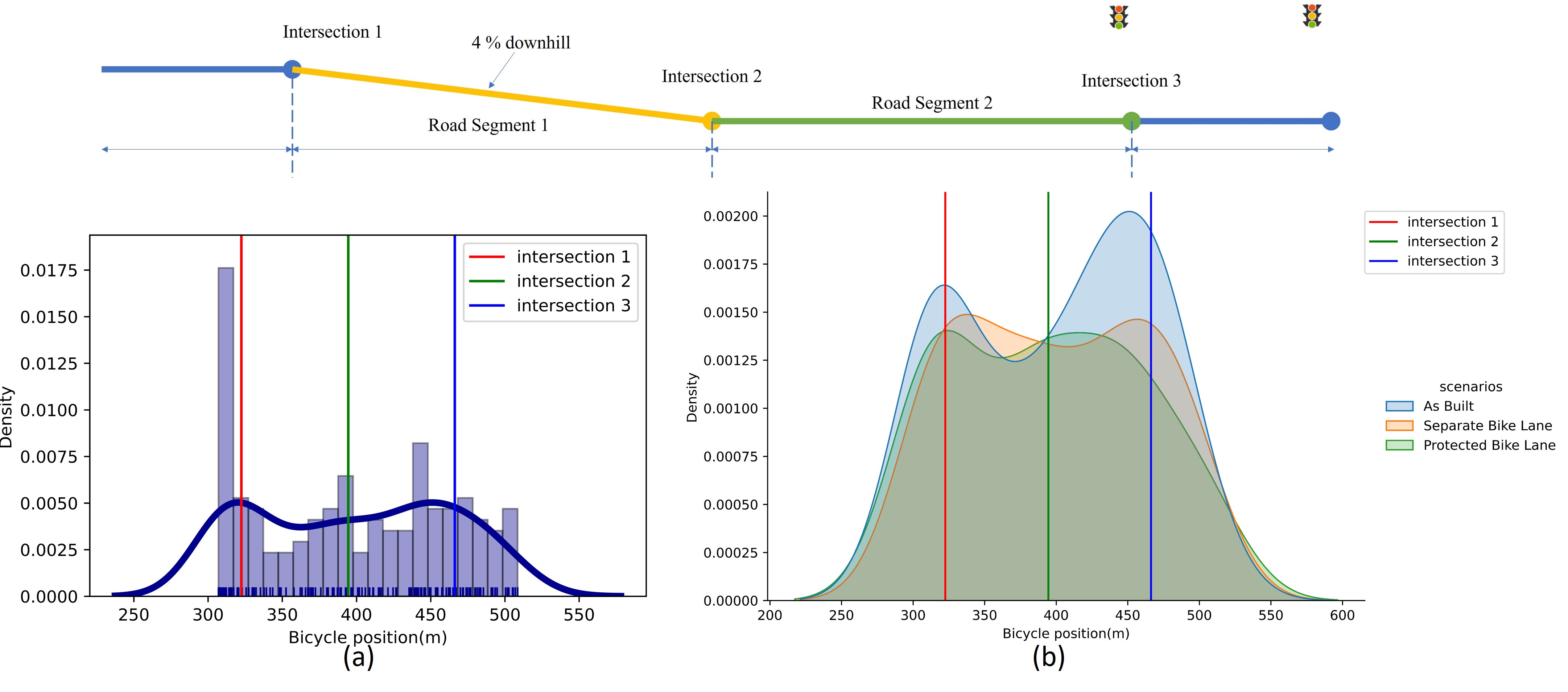}
    \caption{HR change points distribution (a) overall distribution on the road; (b) distribution over each scenario. Note that each scenario appears to have two peaks, with the as-built design having higher peaks in the first intersection and the third intersection. The density plots of separate bike lane and protected bike lane scenarios are smoother than the as-built design, which indicates in the the HR change points in as-built scenario are more subjective to roadway environmental changes. }
    \label{fig:hr_change_point_scenario_distribution}
\end{figure}

In addition to the overall HR level, we are also interested in the abrupt changes in participants' HR. By utilizing the BCP method, we are able to extract the abrupt HR changes for each scenario. Figure \ref{fig:hr_change_point} illustrates the average frequency of HR change points in different scenarios. The LMM model shows that both the separate bike lane ($\beta = -0.393, SE = 0.145, p< 0.01$) and protected bike lane ($\beta = -0.360, SE = 0.145, p< 0.05$) have significantly lower frequency of HR change point than the as-built scenario. The frequency of HR change points in the as-built design is almost twice that of the separate bike lane and protected bike lane. The distribution of HR change points are shown in Figure \ref{fig:hr_change_point_scenario_distribution}. There are three peaks in Figure \ref{fig:hr_change_point_scenario_distribution} - a, where all take place before the participant arrives at an intersection. Among the three intersections, the peak of the HR change point in the third intersection, which has a traffic signal, happens earlier than the other two intersections. Figure \ref{fig:hr_change_point_scenario_distribution} - b is the density plot of HR change points for different scenarios. Each scenario appears to have two peaks, with the as-built design having higher peaks in the first intersection and the third intersection. The density plots of the separate bike lane and protected bike lane scenarios are smoother than the as-built design. This indicates that the HR change points in the as-built scenario are more subjective to roadway environmental changes. In other words, the separate bike lane and protected bike lane may reduce the effect of environment changes to the HR changes.

\section{Discussion}
\subsection{Cycling Performance}
The results show that the roadway design can affect cycling performance. Among the three roadway designs, the speed in the protected bike lane is significantly lower than the other two designs. On average, participants cycled at a lower speed when they are separated from the vehicle lane. However, this contradicts with a similar IVE study (with 50 participants) where bicyclists ride at lower speeds in the no bike lane condition versus the bike lane condition \cite{cobb2021bicyclists}. The differences may be due to (1) the different IVE settings. In \cite{cobb2021bicyclists}, A screen provides the forward view while in this study, a head-mounted display provides the forward view, this implies that different IVE settings will affect bicyclists' responses, as discussed by previous studies \cite{bogacz2020comparison,guo2021orclsim}. (2) the different road environments. Our IVE is modeled from the real-road and all the participants are local bicyclists, they are more familiar with the as-built designs. (3) The vehicle settings. The vehicles are randomly generated based on the empirically observed distribution of vehicle arrivals from the start point with a fixed routine in the vehicle lane. For some participants, when they are cycling in the shared vehicle lane in the as-built scenario, the approaching vehicles from behind will slow down and follow them until the vehicle lane is cleared. Therefore, in those cases, the participants will only see an open street ahead of them without any cars passing by. In the post-experiment survey, some participants mentioned that they were motivated to ride faster when there were no vehicles passing by them. Based on our statistics of how many vehicles have passed the participants during the experiment, the average number of passed vehicle during the experiment for as-built design (0.77) is less than separate bike lane (0.88) and protected bike lane (1.03). (4) The traffic volume and speed. The traffic volume is relatively low in our experiment. On average, there is 0.9 vehicle passing the participant, which is lower than related studies. Water Street is an urban road with a speed limit of 25 mph and vehicles within the IVE were designed to travel constantly at this speed limit, which may indicate that the effect of different roadway designs to bicyclists' speed is subjective to traffic volume and vehicle speed. In addition, some participants report that they feel the protected bike lane is narrower than expected, they want to avoid hitting the pylons during the experiment, which may potentially lower their speed and decrease the lateral distance to the curb. For the lateral lane position, the LMM model result is marginally significant, bicyclists tend to stay away from the vehicle lane when there is a bike lane, which will lead to a larger lateral distance when a vehicle is trying to pass them. This can help to increase bicyclists' comfort level and safety, which has been shown by many previous studies \cite{mcneil2015influence,nazemi2021studying}. 

\subsection{Gaze Behavior}
The gaze behaviors also vary in different roadway designs. Generally, compared with separate bike lane and protected bike lane scenarios, participants in the as-built scenario have a wider horizontal distribution of fixations, indicated by a larger horizontal gaze variability and a lower PRC, which can be a sign of active searching strategy in the environment. The higher PRC in the separate bike lane and protected bike lane scenarios might be an indicator of higher cognitive workload \cite{engstrom2005effects}, but this does not necessarily suggest a sub-optimal state for the gaze strategy. According to the Yerkes-Dodson law \cite{yerkes1908relation}, there is an optimal range of cognitive load; if the current cognitive load is lower than optimal range, increased cognitive workload with higher arousal level can help to improve the performance. The addition of a separate bike lane may have impacted the cognitive workload of the participants to focus on keeping the bicycle in the center of the bike lane. The increase in cognitive workload was lower for protected bike lane, where bicyclists tend to keep closer to the curbside instead of being on the center of the bike lane. 

In the as-built scenario, a shorter fixation duration is observed, which as discussed by previous research is related to a higher hazard estimation of bicyclists \cite{von2020gaze}. It is further verified by our post-experiment survey in which most of the participants rate the as-built scenario as the least safe scenario. In terms of the two alternative designs, the separate bike lane scenario seems to have a more focused gaze behavior than the protected bike lane. The phenomenon is also identified by the gaze entropy results. Only the separate bike lane scenario has a significant lower stationary gaze entropy, which quantifies the overall spatial dispersion of gaze.
To our knowledge, the effect of roadway design on gaze entropy of bicyclists has not previously been examined. Increase in SGE indicates a change in the spatial areas that information is being sampled from, which is illustrated by the less populated and more dispersed depiction of fixation density, as discussed in a driver-related study \cite{shiferaw2018stationary}. 
Meanwhile, for the GTE, no significant differences are found between the three scenarios. The increase in GTE reflects a more random pattern of transitions between fixations. Variation in GTE is related to scene complexity and task demand \cite{shiferaw2019review}. One possible reason for the result may be when designing a separate bike or protected bike lane, participants may feel obligated to maintain lateral bike lane position (especially in IVE), which offsets the effect of new roadway designs with a less scene complexity for necessary visual information retrieval. 

\subsection{Heart Rate Variation}
For the HR response, we did not find any significant difference in mean HR between the three scenarios. The correlations between HR/HRV and subjective safety ratings were also found to be weak in previous naturalistic studies \cite{doorley2015analysis,fitch2020psychological}. In our controlled experiment, the association is even less conclusive. However, when considering the abrupt changes in HR, after extracting the HR change point from the raw data, it is found that both the separate bike lane and protected bike lane scenarios have a significant lower frequency of HR change point than as-built scenario. We further explore the spatial distribution of the HR change point and find that the peaks of the HR change point occur more frequently prior to reaching the intersections. Based on our results, in our low-task requirement scenario, the intersection is more associated with higher number of change points, possibly showing higher stress level than other sites. 

The position of HR change point depends on the types of intersections. The first intersection has more HR change points than the other two intersections. There are two possible reasons. First, as the first intersection is in the beginning of the experiment, participants may still need some time to get used to the IVE, even if they have practiced before. Second, as indicated by previous study, spatially open locations increase the level of perceived risk \cite{von2020gaze}, the road segment after the first intersection is a downhill road. After entering the first intersection, bicyclists will have a wider field of view, which can lead to increased level of perceived risk. Our study further verified this finding by abrupt changes in HR. Moreover, participants seem to have HR change points earlier in more complex intersections (specifically, the third intersection with traffic lights and stop lines). This can be explained by the fact that more complex intersections require more time to prepare for crossing. This aligns with another naturalistic study which shows bicyclists' first fixation to the traffic light occurs earlier in a no bike lane road \cite{rupi2019visual}.  While not in biking studies, similar results were achieved in other transportation studies with respect to other road users such as drivers' stress levels and emotions when getting closer to the intersections. For instance, recent studies both through subjective measures \cite{bustos2021predicting}, self-reports \cite{dittrich2021drivers}, and increases in HR \cite{tavakoli2021harmony,tavakoli2021leveraging} have all shown that drivers experience higher stress level as they arrive at an intersection. However, we note that our findings need to be further verified by future studies due to the limited number and type of intersections in this study. Moreover, the distribution of HR change point for the separate bike lane and protected bike lane scenario are smoother than the as-built scenario. The separate bike lane has more delayed peaks than the other two designs. The reasons behind it should be further explored in future studies. 

One important point with respect to the HR in our study is the short duration of each scenario as well as the overall experiment. Because the HR was sampled at a relatively lower frequency (1 Hz), the number of data points per scenario becomes significantly smaller as compared to other data sources such as gaze measures (120 Hz). The low frequency might be another reason for the insignificant results in the comparison of mean HR across the three scenarios. In the future work of this study, we are planning to use the raw PPG readings from the watches to enhance the depth of the HR modeling within different scenarios. While HR is sampled at 1 Hz, PPG is sampled at 100 Hz but with the caveat of being affected by the motion artifacts. However, we should note that even with a lower number of data points, a change point detector, when applied to the overall data of a participant, can learn the proper distribution and find the moments of abrupt increases, which are spatially intuitive as well (e.g., being close to intersections).

\subsection{Demographics and Survey}  
Although a majority of the 50 participants rate the protected bike lane with pylons scenario as the safest design, the post hoc comparison does not reveal too many differences between the separate bike lane and protected bike lane scenarios. The protected bike lane scenario has a lower average speed. The average lane position is closer to the road curbside and the separate bike lane design has a slightly higher gaze concentration. Other than these findings, there is little evidence showing significant differences between these two alternative designs in terms of cycling behavior and physiological responses. These results indicate there exists some differences between participants' subjective ratings and objective behavioral response. 

No gender or age differences are found to be statistically significant in this study. It is widely accepted that female bicyclists have a lower cycling participation rates \cite{mitra2019can}, as well as stronger preferences than males for greater separation from motor traffic \cite{aldred2017cycling}. While some studies report minor gender differences \cite{cobb2021bicyclists}, it is argued that female bicyclists using the lanes had significantly more positive associations with the protected lanes than males. In other words, the protected bike lanes can somehow help close the gender gap in cycling \cite{dill2014can,aldred2017cycling}. It is worth noting that most of the participants in this study are regular bicyclists, so they are not representative of the entire population. Evidence of stronger preferences among older people is also limited \cite{aldred2017cycling}. Another notable finding from this study is that the realism of the speed is more related to bicyclists' physiological changes than the realism of steering. This can be task-dependent, as in our experiment, most of the road is straight and only a few steering maneuvers are required around the second intersection.

\section{Limitations and Future Work}
Limitations of this study are as follows. First, the duration of the experiment is short, some findings in this experiment needs further study. Building a longer road segment in the IVE with more street blocks can be a solution. However, we should note that longer duration is more likely to cause motion sickness or fatigue, which can lead to performance degradation. Thus, future work needs to find an optimal duration for the study accounting for the trade-off between avoiding motion sickness and retrieving longer time series of data. Second, although a practice scenario is introduced in the beginning and the order of the scenarios is randomized, a learning effect can exist. This means that the participant might become more familiar through the experiments as they progress through the scenarios, which can affect HR, gaze, and speed. Third, more benchmark studies are needed to verify the findings in real-world environments. We have conducted a pilot study of six participants both in IVE and real-road, and preliminary findings show that most of the physiological responses in IVE are representative of the real-world \cite{guo_robartes_angulo_chen_heydarian_2021}. Nonetheless, a benchmark study in the real-world is needed for the same participant groups to validate the findings. Future work will be focused on benchmarking the IVE setup in more diverse scenarios and with a higher number of participants. 

While in this study we focused on the design, it can be the case that different events within each design can impact the participant's psycho-physiology. For instance, when considering the effect of a separate bike lane, it could be the case that increase in traffic density, speed of vehicles, and other environmental factors can affect how participant's psycho-physiology changes within each scenario. In an attempt to isolate the effect of the roadway design, these variables had little or no variation in this experimental design. Future work will be focused on simulating more detailed events within each alternative design to better illustrate the interaction between environmental properties (e.g., presence of bike lane) and events (e.g., vehicles passing with high speed).  

While we focused on HR and gaze, we note that human psycho-physiology is a complicated matter which is not bound by only two measures. The addition of wearable devices with more detailed sensors (e.g., skin conductance, skin temperature, and breathing patterns) may provide additional insight on the bicyclists' psycho-physiology.

\section{Conclusion}
This study explores bicyclists' physiological and behavioral changes in different urban roadway designs. In an immersive virtual environment, a bicycle simulator with integrated mobile sensing devices is used in the experiment to record bicyclists’ behavioral and physiological responses on the same road with three different roadway designs: shared bike lane (as-built), separate bike lane, and protected bike lane with pylons. Results from 50 participants indicate that (1) the protected bike lane design has the highest subjective safety rating; (2) participants in the protected bike lane scenario have the lowest cycling speed and highest lateral distance to the vehicle lane, indicating the potential for safer bicycling behavior with lower speeds and increased separation from vehicles; (3) bicyclists focus their gaze on the cycling task more in the separate and protected bike lane scenarios, indicating the potential for decreased distractions when cycling in a separate or protected bike lane compared to shared bike lane; (4) creating separation zones for bicyclists (whether separate bike lane or protected bike lane) has the potential to reduce the stress level, as indicated by decreased HR changes compared to the shared bike lane; and (5) the immersive virtual environment can be an efficient and safe tool to evaluate bicyclists' behavioral and physiological responses in different alternative roadway designs.

\bibliographystyle{plain}
\bibliography{references}
\end{document}